\documentclass[]{aastex63}
\usepackage{lineno}
\usepackage{amsmath}
\usepackage{mathrsfs}
\usepackage{verbatim}
\usepackage{longtable}
\usepackage{cuted}
\usepackage{appendix}
\usepackage{graphicx}
\usepackage{CJK}

\shorttitle{Jet Structure and Burst Environment of GRB 221009A}
\shortauthors{Ren et al.}
\begin{document}

\title{Jet Structure and Burst Environment of GRB 221009A}

\correspondingauthor{Zi-Gao Dai}
\email{daizg@ustc.edu.cn}

\author[0000-0002-9037-8642]{Jia Ren}
\affiliation{School of Astronomy and Space Science, Nanjing University, Nanjing 210023, China}
\author[0000-0002-8385-7848]{Yun Wang}
\affiliation{Key Laboratory of Dark Matter and Space Astronomy, Purple Mountain Observatory, Chinese Academy of Sciences, Nanjing 210034, China}
\author[0000-0002-7835-8585]{Zi-Gao Dai}
\affiliation{Department of Astronomy, School of Physical Sciences, University of Science and Technology of China, Hefei 230026, China}

\begin{abstract}
We conducted a comprehensive investigation of the brightest-of-all-time GRB 221009A
using new insights from very high energy (VHE) observations from LHAASO
and a complete multiwavelength afterglow dataset.
Through data fitting, we imposed constraints on the jet structure, radiation mechanisms,
and burst environment of GRB 221009A.
Our findings reveal a structured jet morphology characterized by a core+wing configuration.
A smooth transition of energy within the jet takes place between the core and wing,
but with a discontinuity in the bulk Lorentz factor.
The jet structure differs from both the case of short GRB 170817A
and the results of numerical simulations for long-duration bursts.
The VHE emission can be explained by the forward-shock
synchrotron self-Compton radiation of the core component,
but requiring a distinctive transition of the burst environment from uniform to wind-like,
suggesting the presence of complex pre-burst mass ejection processes.
The low-energy multiwavelength afterglow is mainly governed by the synchrotron radiation
from the forward and reverse shocks of the wing component.
Our analysis indicates a magnetization factor of 5 for the wing component.
Additionally, by comparing the forward shock parameters of the core and wing components,
we find a potential correlation between the electron acceleration efficiency
and both the Lorentz factor of the shock and the magnetic field equipartition factor.
We discuss the significance of our findings, potential interpretations, and remaining issues.
\end{abstract}

\section{Introduction}

Gamma-ray bursts (GRBs) are among the brightest stellar-level events in the universe.
The multi-wavelength afterglow radiation of GRBs spans a wide range of the electromagnetic spectrum,
making them valuable subjects and tools in the study of time-domain astronomy.
However, the understanding of the radiation characteristics and origins of GRBs
and their afterglows in the very high energy (VHE; $>100$~GeV) regime
has been limited for a significant period of time because of lack of observations
\citep{Zhang_2001_Meszaros_apj_v559.p110..122,Zou_2009_Fan_mnras_v396.p1163Z,
Piron_2016__ComptesRendusPhysique_v17.p617..631,
Nava_2018__InternationalJournalofModernPhysicsD_v27.p1842003..1842003}.
After the launch of the Fermi detector,
the Large Area Telescope (LAT) recorded a group of GRBs with photon energies $\sim 10$~GeV up to 100~GeV,
which opened a detection window in that energy range
\citep{Nava_2018__InternationalJournalofModernPhysicsD_v27.p1842003..1842003}.
Additionally, detection of the VHE counterpart of GRBs has been a crucial scientific objective and challenge for imaging atmospheric Cherenkov telescopes (IACTs) for a long time
\citep{Acciari_2011_Aliu_apj_v743.p62..62,Aliu_2014_Aune_apjl_v795.p3..3L,
Alfaro_2017_Alvarez_apj_v843.p88..88,Berti_2017_Group__v324.p70..73,Hoischen_2017_Balzer__v301.p636..636}.
As a result, both observation and theory have been driving efforts to expand observations into the VHE window.

In recent years, IACTs have successfully detected the VHE afterglow of several GRBs,
such as GRBs 180720B
\citep[$0.1-0.4$~TeV;][]{Abdalla_2019_Adam_nat_v575.p464..467},
190114C
\citep[$0.3-1$~TeV;][]{MAGICCollaboration_2019_Acciari_nat_v575.p459..463},
190829A
\citep[$0.2-4$~TeV;][]{Collaboration_2021_Abdalla_Science_v372.p1081..1085},
201015A
\citep[$>140$~GeV;][]{Suda-2022-Artero-icrc.confE.797S},
and 201216C
\citep[$70-200$~GeV;][]{Fukami-2022-Berti-icrc.confE.788F,Fukami_2023_Abe__v.p..,Abe_2023_Abe_mnras.tmp.3105A},
thereby extending the detection range to the sub-TeV to TeV energy range.
This achievement represents the final missing element
in the broadband radiation spectrum of GRB afterglows,
signifying the arrival of the TeV era in the field of GRBs
and ushering in an era of new discoveries and cognitive breakthroughs
\citep{Nava_2021__Universe_v7.p503..503,Miceli_2022_Nava_Galaxies_v10.p66..66,
Berti_2022_Carosi_Galaxies_v10.p67..67,Gill_2022_Granot_Galaxies_v10.p74..74}.

The Large High Altitude Air Shower Observatory (LHAASO)
recently reported their observations of the VHE counterpart of GRB 221009A,
claiming the detection of photons with energies exceeding 10~TeV \citep{GCN32677}.
This represents the first instance of humans observing the VHE counterpart of a GRB
in the energy range exceeding 10~TeV \citep{Cao_2023_Aharonian_SciA....9J2778C}.
Additionally, LHAASO is also the first detector to have monitored the VHE afterglow light curve
of a GRB with both extremely high time and energy resolutions.
LHAASO reported their observations and analytical conclusions in the $0.2-7$~TeV energy range
\citep{LHAASOCollaboration_2023_Cao_Science_v380.p1390..1396}.
Thanks to the extremely high brightness of GRB 221009A,
a wealth of afterglow data was collected through multiwavelength observations
\citep[e.g.,][]{Bright_2023_Rhodes_NatureAstronomy_v.p..,
Kann_2023_Agayeva_apjl_v948.p12..12L,Laskar_2023_Alexander_apjl_v946.p23..23L,
LHAASOCollaboration_2023_Cao_Science_v380.p1390..1396,
Ren_2023_Wang_apj_v947.p53..53}.
Combining a rich assortment of multiwavelength afterglow data
together with valuable insights obtained from the newly accessible VHE energy regime,
the study of GRB 221009A has the potential to significantly enhance our understanding of GRB phenomena.
Moreover, it will establish important connections with research domains
associated with, for example, the physics of plasma under extreme conditions,
the stellar formation and evolution, the ultimate phases of massive star lifecycles,
and the environments in which they exist \citep{Miceli_2022_Nava_Galaxies_v10.p66..66}.

In this paper, using a dataset with both long duration time scale and wide energy range,
we conducted an analysis of the radiation mechanism, jet structure, and burst environment of GRB 221009A.
This paper is organized as follows.
In Section~\ref{sec:model}, we introduce our model considerations.
In Section~\ref{sec:method}, we introduce our fitting parameters and method in detail.
The results of model parameter inference and corresponding discussion are presented in Section~\ref{sec:discussion}.
We summarize and conclusion the significance of our results in Section~\ref{sec:summary}.
We take the cosmology parameters as
$H_0=67.8~\rm km~s^{-1}~Mpc^{-1}$, and $\Omega_M=0.308$
\citep{PlanckCollaboration_2016_Ade_aap_v594.p13..13A}.

\section{Model Assumption}\label{sec:model}
Recent LHAASO observations of GRB 221009A have shown
the presence of a remarkably possible early jet break around $T^*+670$~s in the light curve (LC)
at a significance level of $9.2\sigma$, where $T^*=T_0+226$~s after the
the Fermi Gamma-ray Burst Monitor trigger time $T_0$ at 2022~October~9 at 13:16:59~UT
\citep{GCN32636,LHAASOCollaboration_2023_Cao_Science_v380.p1390..1396}.
This compelling finding suggests that
VHE radiation of GRB 221009A emanates from an exceedingly narrow jet.
However, our previous work \citep{Ren_2023_Wang_apj_v947.p53..53},
along with a series of pertinent studies
\citep{Gill_2023_Granot_mnras_v524.p78..83L,Kann_2023_Agayeva_apjl_v948.p12..12L,
Sato_2023_Murase_mnras_v522.p56..60L,OConnor_2023_Troja_ScienceAdvances_v9.p..},
has revealed that the multiband afterglow cannot be satisfactorily
explained by a simple narrow top-hat jet.
Such an oversimplified model leads to untimely fading of afterglow flux
and mismatches in decay indices.
Consequently, the adoption of a structured jet model
becomes indispensable to explicate the data.

\subsection{Jet Structure}
Numerical simulations of relativistic jet propagation through the progenitor star's envelope
\citep[e.g.,][]{Geng_2019_Zhang_apjl_v877.p40..40L,
Gottlieb_2021_Nakar_mnras_v500.p3511..3526,Gottlieb_2022_Moseley_apjl_v933.p2..2L}
have revealed a rich diversity of angular and axial structures in outflowing matter
\citep[e.g.,][]{Salafia_2020_Barbieri_aap_v636.p105..105A}.
These structures are commonly categorized into three distinct components:
the jet, the jet-cocoon, and the ejecta-cocoon.
 Often, structured jets with power-law or Gaussian angular profiles
 are adopted to model these components
\citep[e.g.,][]{Ren_2020_Lin_apj_v901.p26..26L}.
Additionally, in modeling certain GRB afterglow observations,
a commonly used approach is the so-called two-component jet model
\citep[e.g.,][]{Huang_2004_Wu_apj_v605.p300..306,
Filgas_2011_Kruehler_aap_v526.p113..113A}.
In this study, we describe the jet as a tophat inner `core'
encompassed by a power-law structured outer `wing' in the surrounding region.
We consider an axisymmetric structured jet as
\begin{equation}
E_{\rm k,iso}(\theta)=
\left\{
\begin{array}{lr}
E_{\rm k,iso,1}, & \theta < \theta_{j,1}, \\
E_{\rm k,iso,2}\left(1+\frac{\theta}{\theta_{c,2}}\right)^{k_E},
& \theta_{j,1} < \theta < \theta_{j,2},
\end{array}
\right.
\end{equation}
\begin{equation}
\Gamma_0(\theta)=
\left\{
\begin{array}{lr}
\Gamma_{0,1}, & \theta < \theta_{j,1}, \\
\Gamma_{0,2}\left(1+\frac{\theta}{\theta_{c,2}}\right)^{k_{\Gamma}},
& \theta_{j,1} < \theta < \theta_{j,2}.
\end{array}
\right.
\end{equation}
where $E_{\rm k,iso}(\theta)=4\pi\varepsilon(\theta)$
is the isotropic kinetic energy of jet
with $\varepsilon(\theta)$ being the energy per unit solid angle,
$\Gamma_0(\theta)$ is the initial bulk Lorentz factor
and $\theta_{j}$ is the angle of edge of the jet component, respectively.
Subscripts `1' and `2' are used to differentiate between the core and wing.
$\theta_{c,2}$ represents the generalized core opening angle of the wing,
which describes the angle at which the properties of the wing undergo a transition.
We artificially set a minimum value of $1.4$ for $\Gamma_0(\theta)$.
Using the aforementioned formula,
we can mimic various jet structures,
thus inferring possible formation mechanisms for the jet structure of GRB 221009A.

\subsection{Circum-Burst Environment}
The circum-burst environment of GRBs exhibits a complex and diverse nature.
Alongside the conventional environments of free stellar wind and interstellar medium (ISM),
abrupt density transitions in the surrounding medium are often mentioned
\citep{Dai_2002_Lu_apjl_v565.p87..90L}.
Some studies have explained the sudden re-brightening,
fluctuations, and changes in decay slopes in the afterglow LCs
of GRBs using models incorporating density jumps in the medium
\citep[e.g.,][]{Dai_2003_Wu_apjl_v591.p21..24L,
Monfardini_2006_Kobayashi_apj_v648.p1125..1131,
Li_2020_Wang_apj_v900.p176..176}.
Numerical simulations have shown the mass outflows from massive stars
could create a stellar wind bubble within a certain radius
surrounding the star and eventually transition to a uniform ISM
\citep[e.g.,][]{Eldridge_2006_Genet_mnras_v367.p186..200}.

Initial researches did not reach consensus on the environment of GRB~221009A, e.g.,
\cite{Kann_2023_Agayeva_apjl_v948.p12..12L},
\cite{OConnor_2023_Troja_ScienceAdvances_v9.p..}, and
\cite{Sato_2023_Murase_mnras_v522.p56..60L}
 preferred homogeneous,
\cite{Laskar_2023_Alexander_apjl_v946.p23..23L} and
\cite{Ren_2023_Wang_apj_v947.p53..53}
preferred wind-like.
Subsequently, LHAASO reported the detected VHE afterglow
with two distinct ascending slopes of $\propto t^{15}$ and $\propto t^{1.8}$,
and they argue that the slope of $\propto t^{15}$
could be the result of early energy injection.
Furthermore, the observed slope of $\propto t^{1.8}$ deviates from the typical wind environment
but aligns with the scenario of a homogeneous medium surrounding the burst
\citep{LHAASOCollaboration_2023_Cao_Science_v380.p1390..1396},
indicating a discrepancy to the free stellar wind environment implied by the later multiband afterglow observations
\citep{Gill_2023_Granot_mnras_v524.p78..83L}.
In order to reconcile this problem, we propose an `inverted' model
that assumes a homogeneous environment near the progenitor star
and transitions to a wind-like environment at a specific radius $r_{\rm tr}$,
\begin{equation}
n(r)=3\times10^{35}~{\rm cm}^{-1}A_{\star}
\left\{
\begin{array}{lr}
r_{\rm tr}^{-2}, & r\leqslant r_{\rm tr}, \\
r^{-2}, & r>r_{\rm tr},
\end{array}
\right.
\label{medium}
\end{equation}
where $A_{\star}=\dot{M}/10^{-5}M_{\odot}$ yr$^{-1}$
$\left(4\pi v_w/1000~{\rm km~s}^{-1}\right)^{-1}$ is the so-called wind parameter.

\section{The Acquisition and Usage of Data}
\begin{figure*}[htbp]
\centering
\includegraphics[width=0.65\textwidth]{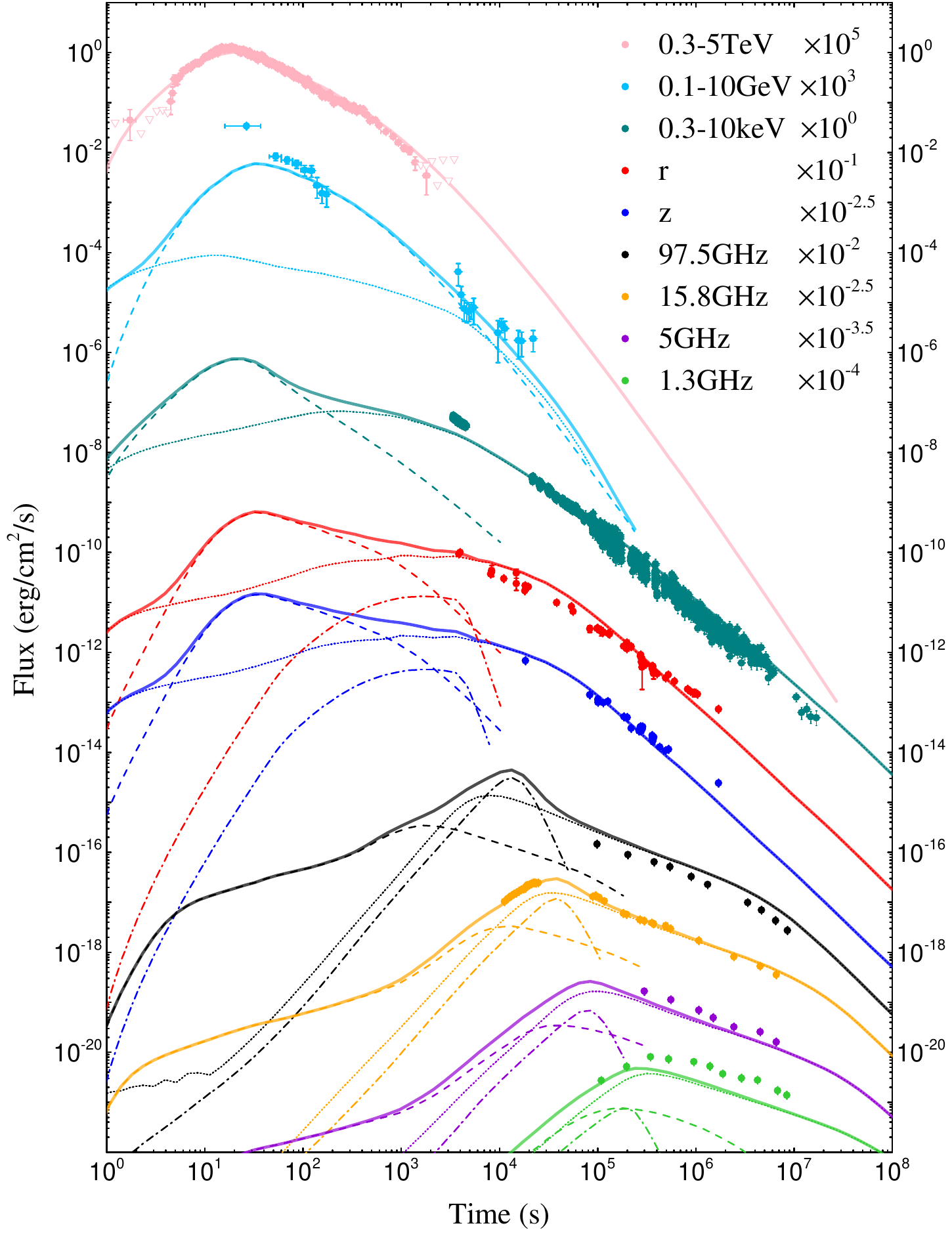}
\caption{Multiband afterglow light curves plotted with best fitting result.
Filled circles are data points and open triangles are upper limits.
Bands are in different colors with shift factors in legend.
Solid lines represent the total emission,
dashed lines represent core forward shock emission,
dotted lines represent wing forward shock emission,
and dash-dotted lines represent wing reverse shock emission, respectively.
}
\label{LCs}
\end{figure*}

\begin{figure*}[htbp]
\centering
\includegraphics[width=0.75\textwidth]{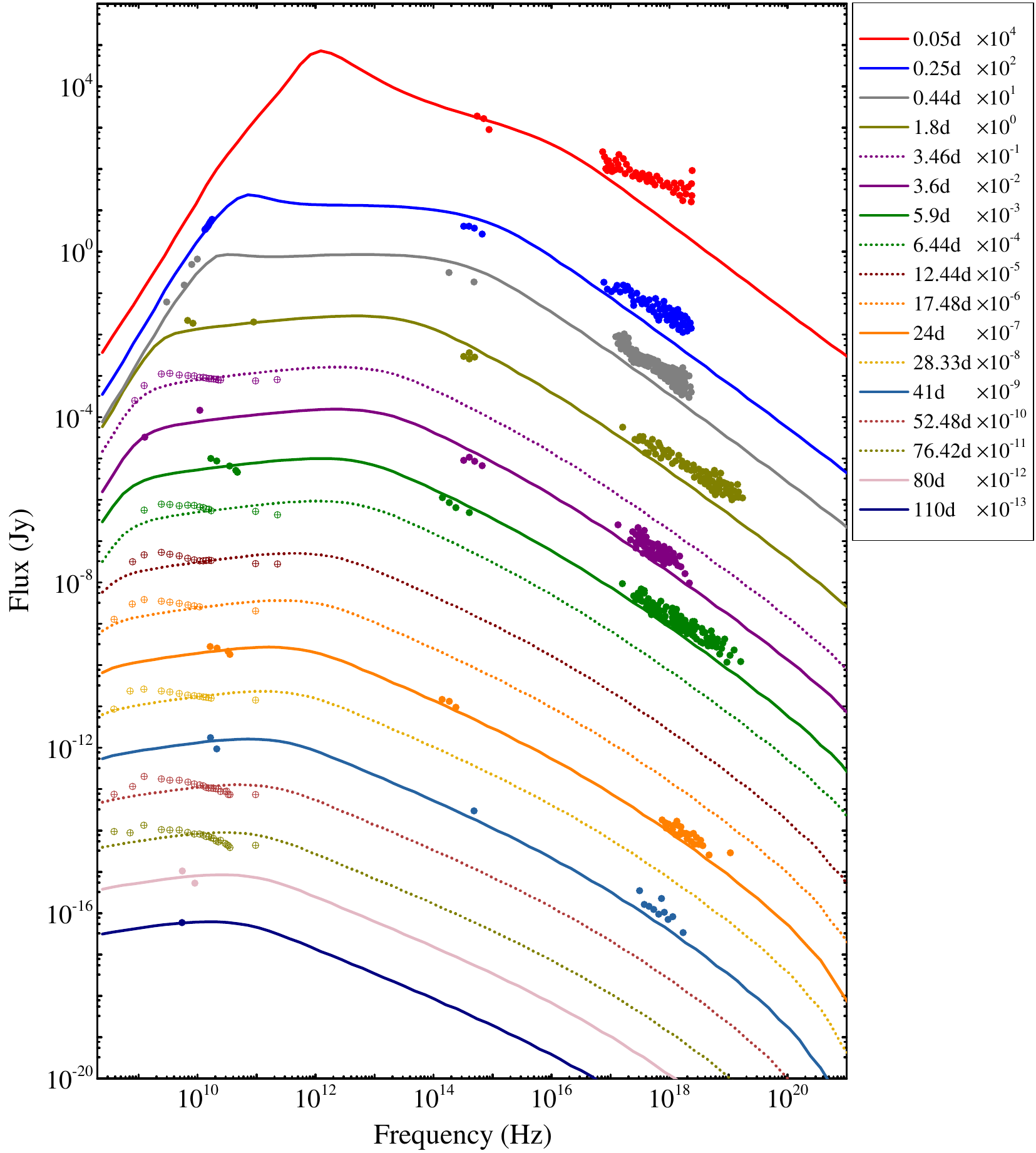}
\caption{SEDs plotted with best fitting result for radio-to-X-ray observations.
Time slices are in different colors with shift factors in legend.
To distinguish between the datasets,
the dashed lines represent the data from \cite{Laskar_2023_Alexander_apjl_v946.p23..23L},
while the solid lines correspond to the data from \cite{OConnor_2023_Troja_ScienceAdvances_v9.p..}.
We interpolated data from \cite{Bright_2023_Rhodes_NatureAstronomy_v.p..} to convert it to 0.25 day.
}
\label{multiday_SEDs}
\end{figure*}

We summarize here the database we used during this work.
We use the VHE LC data and SED data reported in \cite{LHAASOCollaboration_2023_Cao_Science_v380.p1390..1396}\footnote{
\url{https://www.nhepsdc.cn/resource/astro/lhaaso/paper.Science2023.adg9328/}}.
We processed the publicly available data from Fermi-LAT and Swift-XRT and applied them in the fit.
The optical data were obtained from Table~5 column SF11
(i.e., the Galactic extinction is $E(B-V)_{\rm Gal}=1.32$\footnote{
\url{https://ned.ipac.caltech.edu/forms/calculator.html}};
\citealp{Schlafly_2011_Finkbeiner_apj_v737.p103..103}) of
\cite{Kann_2023_Agayeva_apjl_v948.p12..12L}.
We applied host extinction corrections to the data in the $r$ and $z$ bands.
Due to uncertainties in the extinction factors, different articles provide varying values.
We considered an appropriate extinction correction as $E(B-V)_{\rm host}=0.3$ with $R_V=2.93$
\citep{Pei_1992__apj_v395.p130..130} on this basis,
but did not consider the subtraction of late supernova and host galaxy contributions
\citep[e.g.,][]{Blanchard_2023_Villar_arXiv230814197B,Fulton_2023_Smartt_apjl_v946.p22..22L,
Levan_2023_Lamb_apj_v946L..p28L,Shrestha_2023_Sand_apj_v946L..p25S,
Srinivasaragavan_2023_O'Connor_apjl_v949.p39..39S},
so the late optical data may exceed the model curves.
We took radio data from three articles, \cite{Bright_2023_Rhodes_NatureAstronomy_v.p..},
\cite{Laskar_2023_Alexander_apjl_v946.p23..23L}, and \cite{OConnor_2023_Troja_ScienceAdvances_v9.p..}.
The radio LCs used the data from the first two papers (Figure~{\ref{LCs}}),
and radio spectra from all of them (Figure~{\ref{multiday_SEDs}}).

\section{Fitting Method}\label{sec:method}
In order to fit GRB 221009A, we used the {\tt ASGARD} package to generate multiband LCs and spectral energy distributions (SEDs).
Package details are presented in the Appendix~{\ref{sec:appendix}}.
In this work, we have numerically solved both forward shock (FS) and reverse shock (RS) dynamics (see~{\ref{sec:FS_RS}}).
We solve a series of evolution equations independently for the outflow at different jet angles,
which includes the assumption that there is no causality between the elements at different angle,
and leads to the result of structured forward and reverse shock radiation.
The time-dependent electron continuity equation is solved with first-order accuracy
which includes the Compton parameter that is associated with the energy of electrons.
We accounted for synchrotron radiation and synchrotron self-Compton (SSC) emission,
incorporating the corrections due to the Klein-Nishina effect.
Additionally, we considered $\gamma\gamma$ annihilation effects,
the equal-arrival-time surface effect,
and the extragalactic background light (EBL) absorption.
The EBL model we used is taken from
\cite{SaldanaLopez_2021_Dominguez_mnras_v507.p5144..5160},
same as what \cite{LHAASOCollaboration_2023_Cao_Science_v380.p1390..1396} used.

\begin{deluxetable}{ccc}[htbp]
\label{tab1}
\tablecaption{Fitting Result of Model Parameters}
\tablehead{
\colhead{Parameter} & \colhead{Range\tablenotemark{a}} & \colhead{Value\tablenotemark{b}}}
\startdata
$\log_{10}E_{\rm k,iso,1}$ (erg) & $[55,56]$  & $55.32^{+0.16}_{-0.27}$ \\
$\log_{10}\Gamma_{0,1}$    & $[2,3]$  & $2.250^{+0.023}_{-0.039}$ \\
$p_{f,1}$             & $[2,3]$   &  $2.345\pm 0.075$ \\
$\log_{10}\epsilon_{e,f,1}$  & $[-4,-0.1]$   & $-1.18\pm 0.18$ \\
$\log_{10}\epsilon_{B,f,1}$  & $[-8,-1]$   & $-6.03^{+0.24}_{-0.30}$ \\
$\log_{10}\xi_{e,f,1}$         & $[-3,0]$   & $-0.36\pm 0.19$ \\
$\log_{10}\theta_{j,1}$  (rad)    & $[-3,-1]$   & $-2.29\pm 0.11$ \\
\hline
$\log_{10}E_{\rm k,iso,2}$ (erg) & $[53,56.5]$  & $55.69\pm 0.34$ \\
$\log_{10}\Gamma_{0,2}$    & $[1.4,2.5]$ & $1.823^{+0.071}_{-0.099}$ \\
$p_{f,2}$             & $[2,3]$ & $2.145^{+0.020}_{-0.037}$ \\
$\log_{10}\epsilon_{e,f,2}$  & $[-6,-0.1]$ & $-2.34^{+0.29}_{-0.25}$ \\
$\log_{10}\epsilon_{B,f,2}$  & $[-6,-0.1]$ & $-2.95^{+0.24}_{-0.36}$ \\
$\log_{10}\xi_{e,f,2} $        & $[-3,0]$ & $-2.12\pm 0.26$ \\
$\log_{10}\theta_{j,2}$  (rad)    & $[-2,0]$ & $-1.63^{+0.20}_{-0.16}$ \\
\hline
$\log_{10}\epsilon_{e,r,2}$  & $[-6,-0.1]$ & $-2.12^{+1.10}_{-0.91}$ \\
$\log_{10}\epsilon_{B,r,2}$  & $[-6,-0.1]$ & $-4.37\pm 0.87$ \\
$p_{r,2}$           & $[2,3]$ & $2.44^{+0.17}_{-0.26}$ \\
\hline
$\log_{10}\theta_{c,2}$  (rad)    & $[-3,-0.1]$ & $-1.55\pm 0.32$ \\
$k_{E}$           & $[-7,-3]$ & $-5.3^{+1.1}_{-1.3}$ \\
$k_{\Gamma}$      & $[-4,0]$ & $-2.03^{+1.20}_{-0.72}$ \\
\hline
$\log_{10}r_{\rm tr}$   (cm)     & $[15,17]$ & $16.568^{+0.090}_{-0.064}$ \\
$\log_{10}A_{\star}$       & $[-1,1]$ & $0.71^{+0.23}_{-0.14}$
\enddata
\tablenotetext{a}{Uniform prior distribution.}
\tablenotetext{b}{Errors in $1\sigma$.}
\end{deluxetable}

We perform posterior parameter inference using
the Markov Chain Monte Carlo (MCMC) technique,
employing the Python package {\tt emcee} as the sampler
\citep{ForemanMackey_2013_Hogg_pasp_v125.p306..312}.
The sampling was carried out in a 22 dimensional parameter space.
The model parameters include
$\{E_{\rm k,iso,1}$, $\Gamma_{0,1}$, $p_{f,1}$,
$\epsilon_{e,f,1}$, $\epsilon_{B,f,1}$, $\xi_{e,f,1}$,
$\theta_{j,1}$, $E_{\rm k,iso,2}$, $\Gamma_{0,2}$,
$p_{f,2}$, $\epsilon_{e,f,2}$, $\epsilon_{B,f,2}$,
$\xi_{e,f,2}$, $\theta_{j,2}$, $p_{r,2}$,
$\epsilon_{e,r,2}$, $\epsilon_{B,r,2}$,
$\theta_{c,2}$, $k_{E}$, $k_{\Gamma}$,
$r_{\rm tr}$, $A_{\star}\}$,
where the subscription `$f$' and `$r$'
are used to differentiate between FS and RS,
`1' and `2' mark the core and wing component of jet, respectively.

\begin{figure*}[htbp]
\centering
\includegraphics[width=0.95\textwidth]{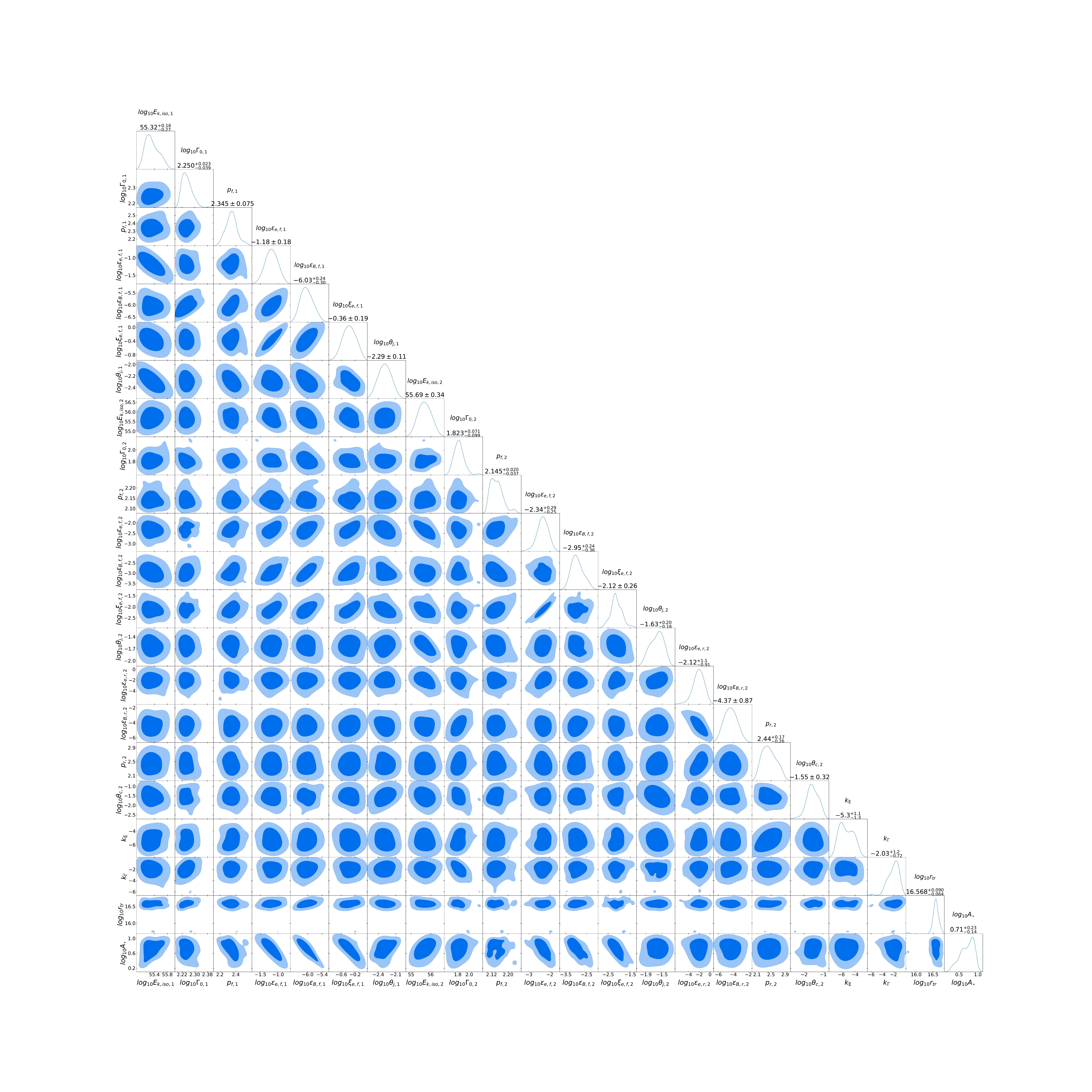}
\caption{Corner plot of the MCMC posterior sample density distributions.}
\label{corner_plot}
\end{figure*}

\begin{figure*}[htbp]
\centering
\includegraphics[width=0.75\textwidth]{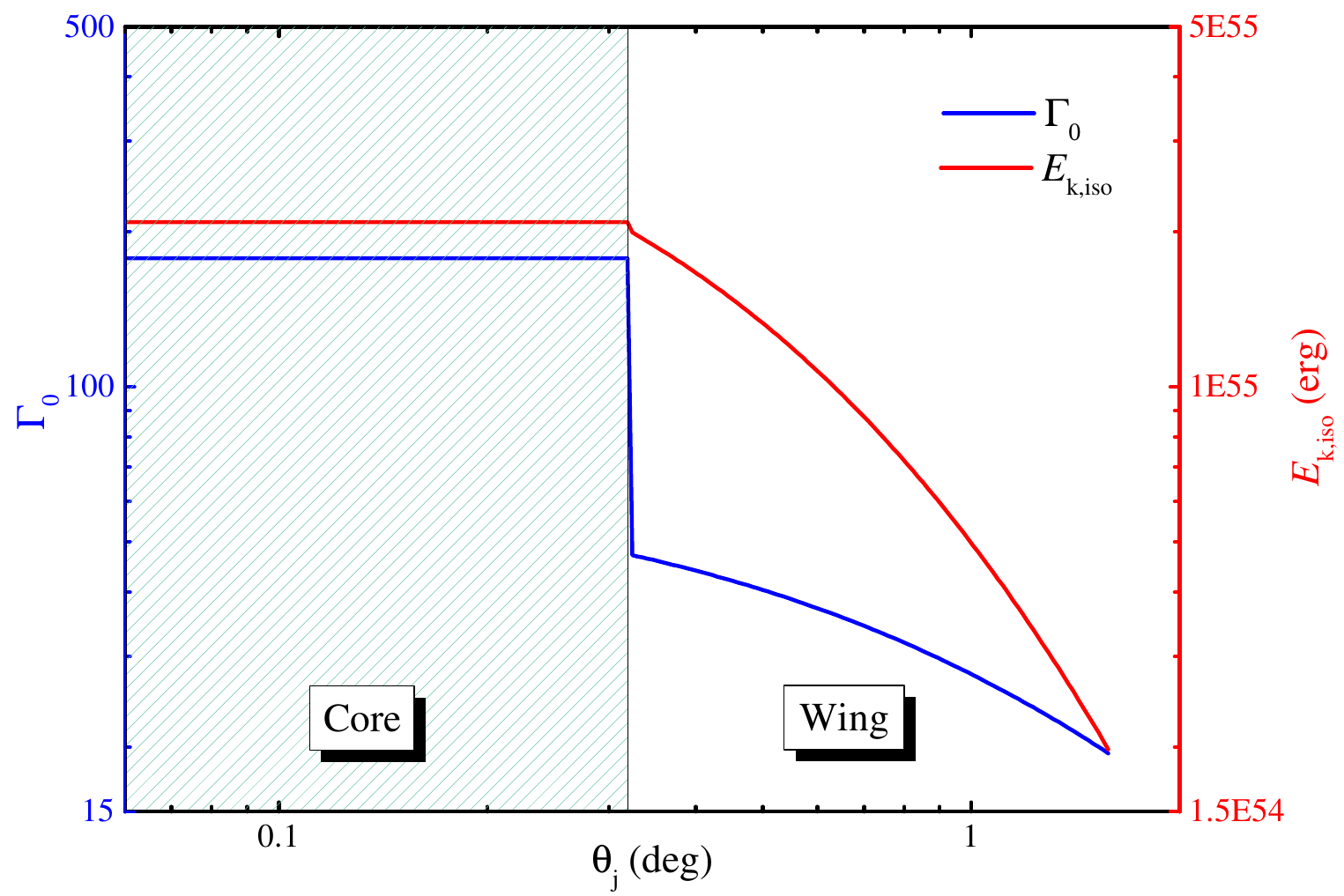}
\caption{Jet structure.}
\label{Jet}
\end{figure*}

\begin{figure*}[htbp]
\centering
\includegraphics[width=0.65\textwidth]{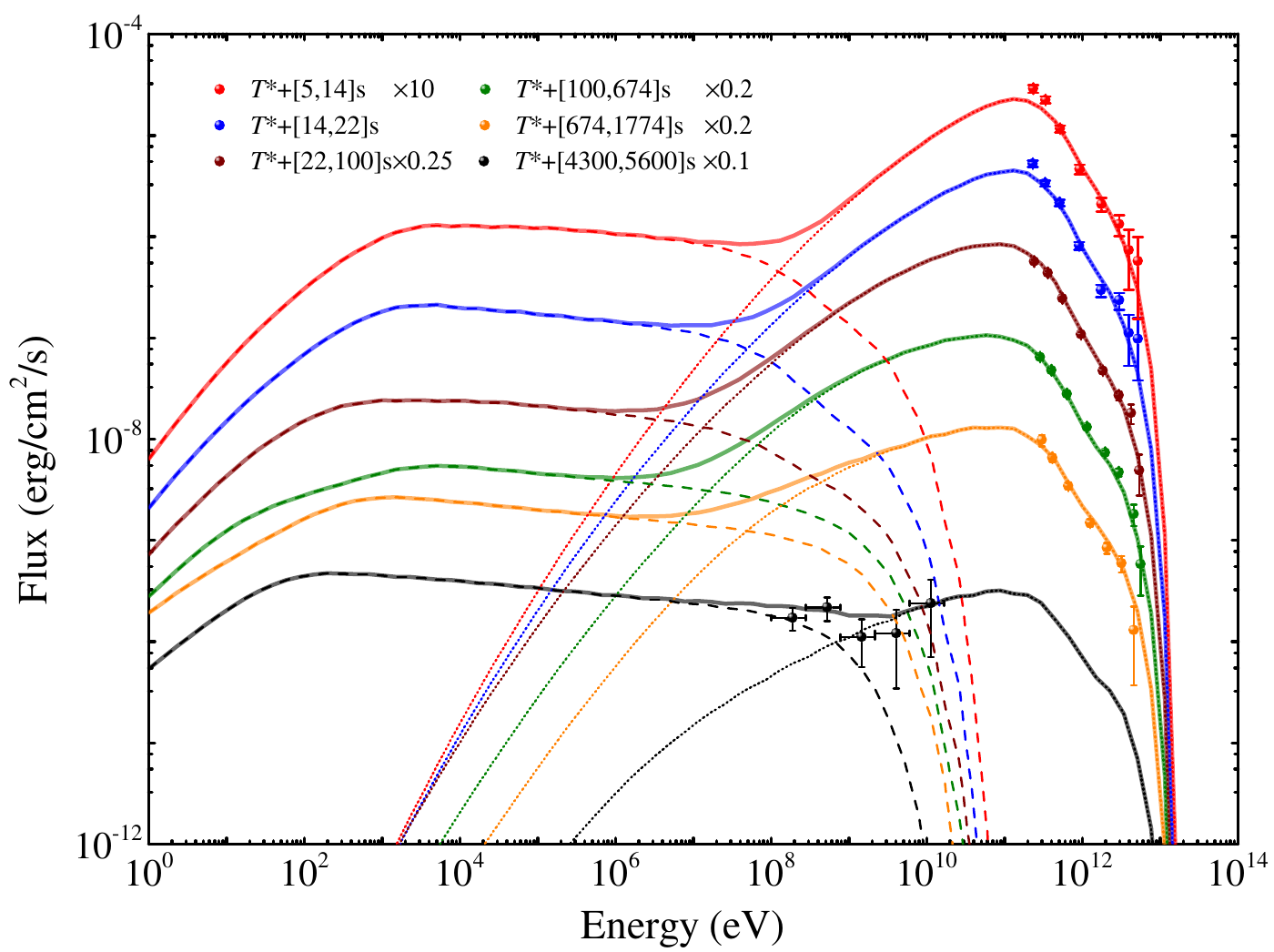}
\caption{SEDs plotted with best fitting result for GeV-to-TeV observations,
data were obtained from \cite{LHAASOCollaboration_2023_Cao_Science_v380.p1390..1396}
and \cite{Ren_2023_Wang_apj_v947.p53..53}.
Time slices are in different colors with shift factors in legend.
Solid lines represent the total emission,
dashed lines represent the synchrotron emission,
and dotted lines represent the SSC emission, respectively.
}
\label{early_SEDs}
\end{figure*}

For simplicity, we have disregarded RS of the core component
and focused solely on the radiation from the FS,
as there is a lack of very early radio-to-optical band observations
to say the RS exist or not.
In contrast, we considered both FS and RS radiation in the wing of the jet,
as early radio observations exhibited signs of RS emission
\citep{Bright_2023_Rhodes_NatureAstronomy_v.p..}.
In our modeling, the electron acceleration efficiency $\xi_{e,r,2}=1$ of RS
is fixed due to the lack of sufficient data to constrain this parameter effectively.

In our fitting process, we simultaneously incorporated both
the radio-to-TeV LCs data and the GeV-to-TeV SEDs data.
This stringent combination provides a highly constraining
posterior distribution for the model parameters.
Our fitting procedure commences with a uniform prior distribution for the parameters.
We adopted a uniform error $\sigma$ of $10\%$ for all observed data points
to ensure equal weighting of the data.
Additionally, we did not account for potential systematic errors
that might arise from differences between instruments\footnote{
The way we do this is oversimplified, and it can lead to underweighting of some well-sampled data.
A more common approach is to introduce a new free parameter in the fitting process
to describe statistical uncertainties and systematic errors in the data
\citep[e.g.,][their Eq. C20]{Salafia_2022_Ravasio_apjl_v931.p19..19L},
and we recommend that future studies consider their fitting strategies more carefully.}.

We sampled the resulting posterior probability density using $50$ walkers in {\tt emcee},
which we ran over 7000 iterations. The resulting marginalized posterior probability density distributions,
after discarding the initial 5000 iterations of the chain as burn-in,
are shown in Figure~\ref{corner_plot}.
The obtained values of parameters are reported in Table~\ref{tab1},
we have categorized the parameters into five groups based on their roles:
core parameters, wing parameters, wing RS parameters,
jet-structure parameters, and environment parameters.

\section{Results and Discussion}\label{sec:discussion}
\subsection{Basic results}
We find that the core component exhibits a typical set of afterglow parameters,
with a large isotropic energy
$E_{\rm k,iso,1}=2.1\times 10^{55}$~erg
and a typical initial bulk Lorentz factor $\Gamma_{0,1}=178$.
Thus the efficiency of prompt emission to the total energy is
$\eta_{\gamma}=42\%$
\citep{An_2023_arXiv230301203A,Yang_2023_Zhao_apjl_v947.p11..11L}.
We calculate the beaming-corrected jet energy
$E_{\rm k,j,1}\approx 2.75\times 10^{50}$~erg for the core,
and total energy $E_{\rm k,j,tot}\approx 1.71\times 10^{51}$~erg for whole jet.
\cite{Fulton_2023_Smartt_apjl_v946.p22..22L} has inferring an explosion energy
of possible exist supernova SN~2022xiw as $E_{\rm k,SN}\approx 3-9\times 10^{52}$~erg.
Also, \cite{Srinivasaragavan_2023_O'Connor_apjl_v949.p39..39S} inferred another possible explosion energy
$E_{\rm k,SN}\approx 1.6-5.2\times 10^{52}$~erg.
Thus, we find the allocation of burst energy is
$\eta= E_{\rm k,j,tot}/(E_{\rm k,j,tot}+E_{\rm k,SN})=1\sim 10 \%$,
which consistent with the founding of \cite{Lue_2018_Lan_apj_v862.p130..130}.
We find that $\Gamma_{0,1}$ deviates slightly from the $2\sigma$ region of
$\Gamma_0-E_{\gamma,\rm iso}$ correlation obtained by
\cite{Liang_2013_Li_apj_v774.p13..13}, but is consistent with those found
by the fittings from prompt emission of \cite{Yang_2023_Zhao_apjl_v947.p11..11L}.
Compared to \cite{LHAASOCollaboration_2023_Cao_Science_v380.p1390..1396},
we find that $p_{f,1}=2.345$ is well consistent with the observed spectral index,
but a smaller half-opening angle $\theta_{j,1}=0.3^{\circ}$ is obtained.
Although such a small half-opening angle presents challenges
for the jet production and collimation process,
we still observe a consistent correlation between the bulk Lorentz factor and the half-opening angle of jet
$\theta_{j,1} \sim 1/\Gamma_{0,1}$,
satisfying the requirements of causality and collimation,
indicating a self-consistent parameter combination.
We find the wing component operates in the mildly relativistic regime with bulk Lorentz factors in the range of tens,
meanwhile the inferred half-opening angle is small, $\theta_{j,2}\sim 1.3^{\circ}$.
We present the inferred angular-dependent isotropic energy $E_{\rm k, iso}(\theta)$
and bulk Lorentz factor $\Gamma_0(\theta)$ distributions of the jet in Figure~{\ref {Jet}}.

\subsection{Jet Structure}
The structure of GRB jets has been a topic of considerable interest.
When studying the afterglow of the short GRB 170817A,
the jet of GRB 170817A has been widely modeled as either Gaussian or power-law profiles,
yielding exciting successes \citep[e.g.,][]{Ren_2020_Lin_apj_v901.p26..26L}.
A thought-provoking question arises as to
whether the jet structure of long and short GRBs
can be consistently described using the same form
\citep{Salafia_2020_Barbieri_aap_v636.p105..105A}.
Although some numerical simulations have shown commonalities between them,
observational constraints on the angular-dependent structure of jet has long been elusive.

Figure~{\ref {Jet}} shows the distribution of isotropic kinetic energy
and bulk Lorentz factor with angle of GRB 221009A.
The findings from $E_{\rm k, iso}(\theta)$ are highly intriguing.
Despite being results obtained through free parameters fitting,
it reveals a remarkably smooth transition between the core and wing regions.
Unlike the energy, the bulk Lorentz factor exhibits a steep transition.
The core region exhibits lower baryon loading,
resulting in faster acceleration and dominating the emission of TeV afterglows (Figure~{\ref{LCs}}).
The wing component operates in the mildly relativistic regime $\Gamma_0\sim 50$ and
gradually transitions to $\Gamma_0 \sim 20$  at larger angles.
This finding suggests a fundamentally different physical origin between the core and wing components.
As shown in Figure~{\ref{Jet}},
the overall angular structure distribution of the jet is inconsistent with a two-component jet.
Instead, this structure is more reminiscent of the properties
exhibited by jet-cocoon systems in numerical simulations
\citep{Gottlieb_2021_Nakar_mnras_v500.p3511..3526,Gottlieb_2022_Moseley_apjl_v933.p2..2L}.
Another evidence is the small wing half-opening angle $\theta_{j,2}\sim 1.3^{\circ}$.
Therefore, in terms of definition, the wing component is closer to a jet-cocoon.
We notice that our study lacks modeling of the ejecta-cocoon with lower velocity,
which may be the reason for inability to explain the low-frequency radio observations satisfactorily
\citep{Laskar_2023_Alexander_apjl_v946.p23..23L}.

\cite{Gottlieb_2021_Nakar_mnras_v500.p3511..3526}
proposes the presence of a jet-cocoon interface region in the angular structure of GRB jets,
where intense mixing processes result in a smooth transition of
jet properties from the core to the wing regions.
We observe similar energy characteristics of GRB 221009A,
but the sharp change in the bulk Lorentz factor reveals
a lack of effective material mixing mechanisms in the interface region,
which contrasts significantly with hydrodynamic simulations.
As a result, our findings may suggest the magnetized nature of the jet,
or, reflect the subsequent impact of the continuous jet from central engine
on the jet structure after the jet-head breaking out of the envelope.

The structure of jet constrained from GRB 221009A differs from that of GRB 170817A.
First, the jet of GRB 170817A has a core with half-opening angle
$\theta_c \sim 3^{\circ}-5^{\circ}$
and a broad wing $\theta_j \sim 20^{\circ}$ \citep[e.g.,][]{Ren_2020_Lin_apj_v901.p26..26L}.
GRB 221009A obviously has smaller values.
Then, the index of decayed energy of GRB 221009A in the wing region
$E(\theta)\propto \theta^{-5.3}$
appear to be consistent with those of GRB 170817A
(e.g., $E(\theta)\propto \theta^{-5.5}$,
\citealp{Ghirlanda_2019_Salafia_Science_v363.p968..971};
$E(\theta)\propto \theta^{-4.5}$,
\citealp{Hotokezaka_2019_Nakar_NatureAstronomy_v3.p940..944};
$E(\theta)\propto \theta^{-6}$,
\citealp{Ren_2020_Lin_apj_v901.p26..26L}).
But, although there is a jump in the bulk Lorentz factor
between the two regions of the jet of GRB 221009A,
the trend of bulk Lorentz factor $\Gamma(\theta)\propto \theta^{-2}$
seems more flat than the jet of GRB 170817A
(e.g., $\Gamma(\theta)\propto \theta^{-3.5}$,
\citealp{Ghirlanda_2019_Salafia_Science_v363.p968..971};
$\Gamma(\theta)\propto \theta^{-4}$,
\citealp{Hotokezaka_2019_Nakar_NatureAstronomy_v3.p940..944};
$\Gamma(\theta)\propto \theta^{-4}$,
\citealp{Ren_2020_Lin_apj_v901.p26..26L}).
It may imply a common physical origin of the structures of both jets,
i.e., the jet passing through a dense material envelope,
but there is a significant difference
in the nature of the envelope that the two jets encounter.
GRB 221009A, being a long GRB,
has its jet passing through a dense and static stellar atmosphere.
In contrast, the jet of short GRB 170817A encounters an expanding
and relatively thin ejecta during the propagation.
As a result, the jet of GRB 221009A undergoes a stronger interaction
with the surrounding material and experiences highly collimation.
Besides, the mixing of material in the cocoon maybe more efficiency,
so that the Lorentz factor more evenly distributed but has lower values.
We believe this explanation could providing a reasonable interpretation
for the unique jet structure observed in GRB 221009A.

Our findings also suggest that a subclass of low-luminosity GRBs
with TeV observations may originate from such jet structures.
The axis of jet deviates slightly from the line-of-sight,
and the observed afterglow could arise from the radiation of a slow-moving jet-cocoon,
as seen in events such as GRBs~190829A and 201015A
\citep{Zhang_2021_Ren_apj_v917.p95..95,Salafia_2022_Ravasio_apjl_v931.p19..19L,
Zhang_2023_Ren_apj_v952.p127..127}.

\subsection{Comparing $\epsilon_e$, $\epsilon_B$, and $\xi_e$ of Core and Wing Components}
For the core component, the derived values of $\epsilon_{e,f,1}=6.6\times 10^{-2}$
and $\epsilon_{B,f,1}=9.3\times 10^{-7}$
are in good agreement with typical values of GRB samples
\citep{BarniolDuran_2014__mnras_v442.p3147..3154,Wang_2015_Zhang_apjs_v219.p9..9,
Beniamini_2017_vanderHorst_mnras_v472.p3161B,Duncan_2023_vanderHorst_mnras_v518.p1522D}.
The small value of $\epsilon_{B,f,1}$
suggests that turbulence behind the shock
amplifies the magnetic field with very low efficiency,
or the turbulence rapidly decays with increasing distance from the shock front
\citep{Lemoine_2013__mnras_v428.p845..866,Lemoine_2015__JournalofPlasmaPhysics_v81.p455810101..455810101}.
We note that the electron acceleration efficiency parameter, $\xi_{e,f,1}=0.44$,
indicates a considerable fraction of electrons
are efficiently accelerated to a relativistic distribution.
This requires the core component of the jet to possess
a powerful acceleration mechanism of particles.

The parameters for the wing component are relatively more intriguing.
Compared to the core, the wing component has a smaller
$\epsilon_{e,f,2}=4.6\times 10^{-3}$,
larger $\epsilon_{B,f,2}=1.1\times 10^{-3}$,
and $\xi_{e,f,2}=7.6\times 10^{-3}$
that is lower by almost two orders of magnitude.
Considering that both components exist in the same surrounding medium
and have similar isotropic kinetic energy of jet,
we can speculate that the lower electron acceleration efficiency
is associated with smaller bulk Lorentz factor and a larger $\epsilon_B$.
Interestingly, TeV-afterglow-detected nearby low-luminosity GRB 190829A
\citep{Collaboration_2021_Abdalla_Science_v372.p1081..1085}
seems to provide evidence for this relevance.
GRB 190829A has a moderate bulk Lorentz factor
$\Gamma_0\sim 45-55$ \citep{Zhang_2021_Ren_apj_v917.p95..95,
Salafia_2022_Ravasio_apjl_v931.p19..19L,Huang_2023_Huang_apj_v947.p84..84}.
As demonstrated by \cite{Salafia_2022_Ravasio_apjl_v931.p19..19L},
it exhibits a very low electron acceleration efficiency
$\xi_{e,f}=7\times 10^{-3}$ (wide prior).
Although there are differences in the fitting results for other parameters,
this relevance may indicate a direction for future research.
Further investigations with a larger sample of well-fitted cases
are expected to verify this correlation.

We also calculate the magnetization in the wing region with
$\mathcal{R}_B\equiv (\xi_{e,r,2}\epsilon_{B,r,2})/(\xi_{e,f,2}\epsilon_{B,f,2})=5$,
notice we need considering the correlation from $\xi_e$
\citep{Eichler_2005_Waxman_apj_v627.p861..867},
which suggests the wing component is weak-magnetized
\citep{Zhang_2003_Kobayashi_apj_v595.p950..954,Japelj_2014_Kopac_apj_v785.p84..84,
Gao_2015_Wang_apj_v810.p160..160,Zhang_2022_Xin_apj_v941.p63..63}.

\subsection{Fitting Results of LCs and SEDs}
In Figure~{\ref{LCs}}, we present the multiband afterglow LCs obtained using the best-fit parameters,
with different line styles distinguishing the emissions from core FS, wing FS, and wing RS, respectively.
Furthermore, in Figure~{\ref{early_SEDs}},
we display the fitted results for the GeV-to-TeV energy range SEDs,
with different line styles representing the core and wing components.
Lastly, Figure~{\ref{multiday_SEDs}} demonstrates the fitted $\nu-F_{\nu}$ spectra for the radio-to-X-ray bands.

The TeV emission is found to be dominated by the core component, as demonstrated in Figure~\ref{LCs}.
By assuming our medium model (Equation~\ref{medium}), all LHAASO observations can be consistently explained.
The rising phase of LC is caused by the uniform ambient medium environment at $r<r_{\rm tr}$.
The rapid decline of LC, observed around $T^*+700$~s,
is attributed to a smooth jet break occurring within the free stellar wind environment.

The broadband emission, spanning from radio to X-ray,
is primarily determined by the radiation from both FS and RS of the wing component.
Observations in the optical and X-ray bands are dominated by FS radiation.
Some detectors indicate the presence of persistent power-law components in the spectra of tailed prompt emission.
The inferred X-ray LC is extend from $T^*+100$ to $T^*+2000$~s, showing a transition from flat to steep decay.
This radiation component is attributed to external shock emission in some studies
\citep{An_2023_arXiv230301203A,Lesage_2023_Veres_apjl_v952.p42..42L,
Zhang_2023_Huang_apj_v956L..p21Z}.
However, we suggest that this early X-ray LC cannot be explained by
external shock radiation because a satisfactory simultaneous fit cannot be reached
for both the radio and X-ray bands.
Instead, we are inclined to attribute this radiation to the `late prompt emission'.
In other words, the continuous and smooth emission characterized by
a power-law decay over time is produced by the internal dissipation of shells
from late-time central engine activity. Meanwhile, the transition in the decay slope
can also be explained self-consistently within this model
\citep{Ghisellini_2007_Ghirlanda_apjl_v658.p75..78L}.

The early-time radio observations show a good fit with both FS and RS radiation,
whereas RS radiation dominates the late-time radio observations.
However, we still lack a precise explanation for the late-time radio observations.
Specifically, the low-frequency flux is significantly underestimated (see Figure~\ref{multiday_SEDs}).
It is possible that we have overlooked an additional trans-relativistic outflow component
known as the ejecta-cocoon, given our employed modeling strategy.
Introducing this component into the model could potentially lead to an improvement in the fitting.
However, considering the satisfactory results achieved with our current model,
which already includes up to 22 parameters,
it may not be worthwhile to add several additional parameters to describe the properties of the ejecta-cocoon.
It would be more advantageous to leave this aspect of modeling for future investigations,
as continuous radio observations will provide late-time data.
Hence, we only briefly discuss this possibility here.

Another potential explanation for the late-time radio emission is the reheating
of low-energy electrons through the synchrotron self-absorption (SSA) process.
This results in the accumulation of heated electrons around the self-absorption frequency,
leading to a noticeable bump in the SEDs near this frequency,
as demonstrated in previous studies
\citep[e.g.,][]{Yang_2016_Zhang_apjl_v819.p12..12L,Li_2020_Yang_apj_v896.p71..71}.
Additionally, the investigation of thermal electron radiation
presents an interesting direction, as it can also have an impact on the radio band \citep{Eichler_2005_Waxman_apj_v627.p861..867,Fukushima_2017_To_apj_v844.p92..92,
Margalit_2021_Quataert_apjl_v923.p14..14L,Warren_2022_Dainotti_apj_v924.p40..40,
Nedora_2023_Dietrich_mnras_v520.p2727..2746}.
However, both SSA and thermal electron radiation are only significant during the early stages of afterglow.
Consequently, their effectiveness in explaining the late-time low-frequency radio emission might be limited.

During the investigation, we also observed that the behavior of GRB 160625B
at low frequencies during the afterglow phase is similar to that of GRB 221009A
\citep{Alexander_2017_Laskar_apj_v848.p69..69,Cunningham_2020_Cenko_apj_v904.p166..166,
Kangas_2020_Fruchter_apj_v894.p43..43}.
Notably, at 12 and 22 days, GRB 160625B exhibited a re-emergence of low-frequency radio emission.
This phenomenon cannot be explained by the RS mechanism
and is consequently attributed to an extreme scattering event (ESE)
\citep{Alexander_2017_Laskar_apj_v848.p69..69}.
In contrast to GRB 160625B, GRB 221009A has a low Galactic latitude.
This factor increases the probability of encountering dense plasma structures
that function as lenses along the path of photon propagation \citep{Tiengo_2023_Pintore_apjl_v946.p30..30L,Vasilopoulos_2023_Karavola_mnras_v521.p1590..1600}.
Consequently, GRB 221009A might be another example of an ESE.

\subsection{Circum-Burst Environment}
The environment surrounding long GRBs has garnered considerable research interest
\citep{Mirabal_2003_Halpern_apj_v595.p935..949,Chevalier_2004_Li_apj_v606.p369..380,
RamirezRuiz_2005_GarciaSeg_apj_v631.p435..445,Eldridge_2006_Genet_mnras_v367.p186..200}.
Although observations have confirmed the association of long GRBs with supernovae,
indicating their connection to the collapse of massive stars
\citep[e.g.,][]{Woosley_2006_Bloom_araa_v44.p507..556},
studies of long GRB afterglow samples reveal that a significant portion of afterglows
does not exhibit characteristics of a free stellar wind environment
\citep[e.g.,][]{Schulze_2011_Klose_aap_v526.p23..23A,Liang_2013_Li_apj_v774.p13..13,
Yi_2013_Wu_apj_v776.p120..120,Yi_2020_Wu_apj_v895.p94..94,
Tian_2022_Qin_apj_v925.p54..54}.
This complexity suggests the diverse origins of the environments surrounding the bursts.
Intriguingly, evidence of one or multiple dense material shells surrounding the progenitor star
has been discovered in certain supernovae
\citep[e.g.,][]{Corsi_2014_Ofek_apj_v782.p42..42,Milisavljevic_2015_Margutti_apj_v815.p120..120}
and interacting superluminous supernovae
\citep[e.g.,][]{Liu_2018_Wang_apj_v856.p59..59}.
The density distribution of these shells often suggests a homogeneous medium,
as opposed to a free stellar wind medium.
In summary, these findings imply complex and unstable material ejection behavior
during the post-main-sequence lifetime of massive stars
\citep{Langer_2012__araa_v50.p107..164}.

This work utilizes a phenomenological model
described by Equation~{\ref{medium}} to tackle the challenge of
explaining the VHE afterglow observations in GRB 221009A.
By considering the transition from a homogeneous
to a wind-like medium at $r_{\rm tr}=3.7\times 10^{16}$~cm,
we have achieved a successful fit to the data.
The number density of medium within
$r \leqslant r_{\rm tr}$ is $n_{\rm tr}=1125$~cm$^{-3}$.
It should be noted that the VHE LC exhibits a rapid rise at approximately $T_b \sim T^*+5$~s.
In our proposed scenario, the rapid rise can be attributed to
the initiation of an external shock when the jet collides with the shell located at $r_{\rm s}$.
This collision implies a lower density at $r < r_{\rm s}$ compared to $n_{\rm tr}$.
Based on this assumption, we can estimate the position of the shock front at this time to be
$r_{\rm s}=2\Gamma_{0,1}^2cT_b/(1+z)=8.2\times 10^{15}$~cm.
In this context, some TeV photons detected prior to $T^*+5$~s,
surpassing the background level,
can be attributed to either internal shock events or collisions
between the jet and material clumps or shells located at $r < r_{\rm s}$
\citep{Marchenko_2007_Foellmi_apjl_v656.p77..80L,Cantiello_2009_Langer_aap_v499.p279..290}.
The lower energy gamma-rays may be obscured by the prompt emission during this period.

Our modeling of the VHE afterglow of GRB 221009A
suggests the presence of a homogeneous shell between $r_{\rm s}$ and $r_{\rm tr}$.
Assuming a spherically symmetric distribution of the shell,
we calculate its mass to be approximately $2\times 10^{-4}~M_{\odot}$,
which is smaller compared to the shell masses observed in supernovae and superluminous supernovae.
In addition, if we assume a stable free stellar wind,
this shell cannot be attributed to the jet shell of the precursor of GRB 221009A.
If it were, it would gather considerably more material than our calculations show, contradicting our results.

We propose that this shell reflects the evolution of material outflow
just prior to the death of the star.
A sudden decline in the strength of the stellar wind,
potentially accompanied by a minor ejection event,
may explain the presence of this shell.
Taking into account the assumption that the progenitor is a typical Wolf-Rayet star,
the velocity of the ejection shell is often slower than the stellar wind velocity
\citep{Milisavljevic_2015_Margutti_apj_v815.p120..120}.
Consequently, the structure of the ejection shell becomes dispersed during its propagation,
thus preventing the formation of shocks in the free stellar wind.
This is a plausible explanation for the smooth transition in the density of the medium.

\section{Summary and Conclusion}\label{sec:summary}
GRB 221009A, as a powerful GRB event,
has provided a rich and complete set of GRB afterglow observations,
offering a comprehensive showing stage for the decades-developed theories
in the field of GRB afterglow.
Our study showcases the success of theoretical efforts,
but also show some outstanding questions in future researches.
\begin{itemize}
\item Our study suggests that GRB 221009A possesses a core+wing structure,
where $E_{\rm k,iso}(\theta)$ shows a smooth transition,
but $\Gamma_0(\theta)$ displays a clear interface stratification between the core and wing.
We neglected the ejecta-cocoon component in our modeling,
but the underestimation of late-time low-frequency radio flux suggests its possible existence,
and future observations of GRB 221009A will help answer this question
\citep{Bright_2023_Rhodes_NatureAstronomy_v.p..,Laskar_2023_Alexander_apjl_v946.p23..23L}.
The structure of the jet in GRB 221009A is differ from that of short GRB 170817A,
possibly due to differences in the jet magnetization and the envelope properties of progenitor.
Our findings also suggest that a subclass of low-luminosity GRBs with VHE observations,
e.g., GRBs 190829A and 201015A, may originate from such jet structures.

\item The fittings demonstrates that SSC emission from core component
can successfully explain the LHAASO observations,
while the lower energy afterglow requires the presence of the synchrotron emission
from wing component for explanation.
However, the previously reported $>10$~TeV photons
clearly cannot be explained by the SSC process \citep{GCN32677}.
Further investigation is needed to determine whether observable hadronic processes
are present during the afterglow phase
\citep[e.g.,][]{Sahu_2022_Valadez_ApJ...929...70S,Isravel_2023_Begue__v.p..}.

\item Our fitting reveal a distinct transition in the burst environment
from uniform to wind-like medium,
indicating the presence of intricate pre-outburst mass ejection processes.
The detection by LHAASO of the early VHE afterglow provides a novel avenue
for investigating the burst environment of progenitor star.
More observations of  VHE-afterglow-detected GRBs will help to
unveil more details about the burst environment.

\item  By comparing the FS parameters of the core and wing components,
we suggest a possible correlation between electron acceleration efficiency $\xi_e$
and both the Lorentz factor of the shock and the magnetic field equipartition factor $\epsilon_B$.
To test this hypothesis,
analyses require using a set of well-sampled afterglow data of GRBs,
and performing plasma numerical simulations under extreme conditions.

\item The complex radio and millimeter emission are difficult to explain within
the scope of standard synchrotron emission theory \citep{Bright_2023_Rhodes_NatureAstronomy_v.p..,
Laskar_2023_Alexander_apjl_v946.p23..23L,
OConnor_2023_Troja_ScienceAdvances_v9.p..}.
Although we did not discuss the aspect of evolving microphysical parameters,
$\epsilon_e$, $\epsilon_B$, $\xi_e$, and $p$, in our work,
considering such evolution may potentially improve the current fitting results
\citep[e.g.,][]{Yang_2018_Zou_ResearchinAstronomyandAstrophysics_v18.p18..18}.

\end{itemize}

\appendix
\section{Model Details}
\label{sec:appendix}
\twocolumngrid
\restartappendixnumbering

This section describes the modeling details of the code package utilized for fitting.

\subsection{Code Framework of \tt{ASGARD}}
A Standard Gamma-ray Burst Afterglow Radiation Diagnoser ({\tt ASGARD}) is a software package
designed to generate multiband afterglow lightcurves (LCs) and spectral energy distributions (SEDs) for specific events.
The package provides a Python interface with underlying FORTRAN code
that allows users to easily conduct research after a simple learning and setup process.

To ensure code reusability, extensibility, and readability,
{\tt ASGARD} follows a modular approach where different modules handle specific calculations.
Python code is used to call these modules.
{\tt ASGARD} currently encompasses three main components:
jet dynamics, electron distributions, and radiation physics.
Radiation physics includes synchrotron radiation and synchrotron self-Compton (SSC) radiation of electrons.

All the modeling in {\tt ASGARD} is based on detailed numerical calculations,
aiming to realistically describe the radiation behavior of the afterglow.
While {\tt ASGARD} follows the `standard' fireball model,
it also incorporates post-standard models developed over decades,
such as energy injection and structured jets.
The extensible architecture of the package allows for the inclusion of additional considerations in the future.
A brief description of the numerical methods and models used is provided in the subsequent subsections.

\subsection{Dynamics of Jet propagation}
During the expansion of a relativistic fireball in a surrounding medium,
two shock waves are created: a forward shock (FS) and a reverse shock (RS).
Upon the formation of shocks, a region of shock interaction arises,
characterized by two distinct components: the shocked medium and the shocked shell.
These two regions are separated by the contact discontinuity surface.
The evolution of these shock waves and the physical properties of the fireball
involves solving a set of equations that couple relativistic (magneto)hydrodynamic waves.
\cite{Blandford_1976_McKee_PhysicsofFluids_v19.p1130..1138}
derived a self-similar solution for adiabatic relativistic hydrodynamic waves,
which provides insights into the time-dependent evolution of the fireball shocks.
However, to obtain a more precise description of the evolution of jet,
a set of differential equations incorporating radiation cooling in one dimension is employed.
These equations account for the energy lost through radiation as the fireball interacts with its surroundings.

The solution of these equations relies on the assumption of the single-shell hypothesis
where the majority of the mass and energy of the fireball
are concentrated within a small radius in the vicinity of the shock region.
This assumption simplifies the calculations and allows for a focused analysis
of the shock dynamics and radiation behavior.

\subsubsection{Dynamics of coupled forward-reverse shock}
\label{sec:FS_RS}
At the beginning of external shock generation,
RS propagates into the jet shell, and FS/RS dynamics are coupled at this stage.
There is a possibility that the RS strength may be suppressed if the jet is magnetized
\citep[e.g.,][]{Zhang_2005_Kobayashi_apj_v628.p315..334,Chen_2021_Liu_mnras_v504.p1759..1771}.
At present, {\tt ASGARD} does not include magnetizing jet modeling.
As a substitute, we used model of
\cite{Yan_2007_Wei_ChineseJournalofAstronomyandAstrophysics_v7.p777..788}
to analyse RS dynamics in a pure fireball scenario.

A homogeneous cold shell is expelled from the central engine,
carrying an isotropic energy $E_{\rm k,iso}$, and possesses a width $\Delta_0$.
As it expands relativistically with a Lorentz factor $\Gamma_0$,
the shell mass is given by $m_0 = E_{\rm k,iso}/\Gamma_0 c^2$, where $c$ is the speed of light.

Within the shocked region, it is assumed that the bulk velocity and energy density are homogeneous,
implying that $\gamma_2 = \gamma_3$ and $e_2 = e_3$.
Hereafter, the subscript `2' designates the shocked medium,
the subscript `3' corresponds to the shocked jet shell,
and the subscript `4' pertains to the unshocked jet shell,
i.e., $\gamma_4=\Gamma_0$.

The comoving number density of the ejecta in region 4 is
$n_{4}=m_0 / (4 \pi r^2 \Delta m_p \gamma_4)$, with $\Delta=\max \left(\Delta_{0}, r / \gamma_4^2\right)$,
which is the width of shell after allowing for spreading.
The swept-up mass of RS is
\begin{equation}
\frac{dm_3}{dr}=4 \pi r^2\left(\beta_4 -\beta_{\rm RS}\right)\gamma_4 n_4 m_p,
\end{equation}
where $\beta = \sqrt{1-1/\gamma^2}$, and
\begin{equation}
\beta_{\rm RS}=\frac{\gamma_3 n_3 \beta_3-\gamma_4 n_4 \beta_4}{\gamma_3 n_3-\gamma_4 n_4}.
\end{equation}

The overall dynamical evolution of RS and FS can be obtained by solving
\begin{equation}
\frac{d\gamma_2}{dr}=-4 \pi r^2 \frac{W}{P},
\end{equation}
where
\begin{equation}
W= \left(\gamma_2^2-1\right) n_1 m_p+\left(\gamma_2 \gamma_{34}-\gamma_4\right)
\left(\gamma_4 n_4 m_p\right)\left(\beta_4-\beta_{\rm RS}\right),
\end{equation}
and
\begin{equation}\label{eq:RS}
\begin{array}{ll}
P & = m_2+m_3 \\
&+\left(1-\epsilon_2\right)\left(2 \gamma_2-1\right)m_2
+\left(1-\epsilon_3\right)\left(\gamma_{34}-1\right)m_3 \\
& +\left(1-\epsilon_3\right) \gamma_2 m_3\left[\gamma_4\left(1-\beta_2 \beta_4\right)
-\frac{\eta \beta_4}{\gamma_2^2 \beta_2}\right].
\end{array}
\end{equation}
Here,
$\gamma_{34}=\gamma_3\gamma_4(1-\beta_3\beta_4)$,
$\epsilon_2$ and $\epsilon_3$ will be described in Section~\ref{sec:electron}.

\subsubsection{Dynamics of forward shock}
Considering only the existence of FS,
a series of papers give their results.
We quote the dynamics of FS ($\Gamma=\gamma_2$) described as
\citep{Nava_2013_Sironi_mnras_v433.p2107..2121,Zhang_2018pgrb.book.....Z},
\begin{equation}\label{eq:dGamma}
\resizebox{0.5\textwidth}{!}{$
\dfrac{d\Gamma}{dr}=
-\dfrac{\Gamma(\Gamma^2-1)(\hat{\gamma}\Gamma-\hat{\gamma}+1)\frac{dm}{dr}c^2
-(\hat{\gamma}-1)\Gamma(\hat{\gamma}\Gamma^2-\hat{\gamma}+1)(3U/r)}
{\Gamma^2[m_0+m]c^2+(\hat{\gamma}^2\Gamma^2-\hat{\gamma}^2+3\hat{\gamma}-2)U}
$},
\end{equation}
\begin{equation}\label{eq:diff_U}
\frac{dU}{dr}=(1-\epsilon)(\Gamma-1) c^{2}\frac{dm}{dr}
-(\hat{\gamma}-1)\left(\frac{3}{r}-\frac{1}{\Gamma} \frac{d \Gamma}{d r}\right) U,
\end{equation}
where ${dm}/{dr}=4 \pi r^{2} n(r) m_p$
with $n(r)$ being the particle density of
circum-burst medium and $m_p$ being the proton mass,
and $\Gamma(r)$, $m(r)$, $U(r)$, and $\epsilon$
are the bulk Lorentz factor, the swept-up mass,
the internal energy, and the radiation efficiency
of electrons in the external-forward shock, respectively.
The radiation efficiency has affected by radiation processes
and will be describe in next section.
The adiabatic index is $\hat{\gamma} \simeq
(5-1.21937\zeta+0.18203 \zeta^{2}
-0.96583 \zeta^{3}+2.32513 \zeta^{4}
-2.39332 \zeta^{5}+1.07136 \zeta^{6})/3$
with $\zeta \equiv \Theta /(0.24+\Theta)$,
$\Theta \simeq
(\Gamma \beta/3)
\left[\Gamma \beta+1.07(\Gamma \beta)^{2}\right]/
\left[1+\Gamma \beta+1.07(\Gamma \beta)^{2}\right]$
(\citealp{PeEr_2012__apjl_v752.p8..11}).

\subsection{Solving of the Electron Continuity Equation}\label{sec:electron}
We denote the instantaneous electron distribution as ${dN_e}/d\gamma_e^{\prime}$,
of which the evolution can be solved based on the continuity equation of electrons
\citep{Fan_2008_Piran_mnras_v384.p1483..1501},
\begin{equation}
\frac{\partial}{\partial r}\left(\frac{dN_e}{d\gamma_e^{\prime}}\right)
+\frac{\partial}{\partial\gamma_e^{\prime}}
\left[\frac{d\gamma^{\prime}_e}{dr}\left(\frac{dN_e}{d\gamma^{\prime}_e}\right)\right]
= Q\left(\gamma^{\prime}_e,r\right)~,
\end{equation}
where ``$\prime$" marks the co-moving frame of shock, and
\begin{equation}
\frac{d\gamma^{\prime}_e}{dr}=-\frac{\sigma_{\rm T}}{6 \pi m_e c^2}
\frac{B^{\prime 2}}{\beta \Gamma}\left[1+Y\left(\gamma^{\prime}_e\right)\right]
{\gamma^{\prime}_e}^2-\frac{\gamma^{\prime}_e}{r}
\end{equation}
is the cooling term which has included
the cooling effects of synchrotron radiation,
SSC process, and adiabatic expansion.
The swept-in electrons by the shock are accelerated
to a power-law distribution of Lorentz factor $\gamma^{\prime}_e$,
i.e., $Q\propto {\gamma^{\prime}_e}^{-p}$ for
$\gamma^{\prime}_m
\leqslant \gamma^{\prime}_e \leqslant \gamma^{\prime}_M$,
where $p (>2)$ is the power-law index.
Here,
$\gamma^{\prime}_m=\epsilon_e/\xi_e(p-2)m_p\Gamma/[(p-1)m_e]+1$
(\citealp{Sari_1998_Piran_apj_v497.p17..20L}),
with $\Gamma=\gamma_2$ for FS, and $\Gamma=\gamma_{34}$ for RS, respectively.
We assume that a fraction $\xi_e$ of the electron population is shock-accelerated
\citep{Eichler_2005_Waxman_apj_v627.p861..867}.
Evidently, the overall inflow of nonthermal electrons should be
\begin{equation}\label{eq:Qinj}
\int Q\left(\gamma^{\prime}_e,r\right) d\gamma^{\prime}_e
=4 \pi r^2  n(r) \xi_e dr.
\end{equation}
The maximum Lorentz factor of electron is
$\gamma^{\prime}_M=\sqrt{9m_e^2{c^4}/[8B'{q_e}^3(1+Y)]}$
with $q_e$ being the electron charge
\citep{Kumar_2012_Hernandez_mnras_v427.p40..44L},
where $Y$ is the Compton parameter.
We note that the Compton parameter
\begin{equation}\label{eq:Y}
Y(\gamma_e^{\prime})=\frac{-1+\sqrt{1+4 \epsilon_{\rm rad} \eta_{\rm KN} \epsilon_e / \epsilon_B}}{2}
\end{equation}
has been solved based on the work of \cite{Fan_2006_Piran_mnras_v369.p197..206} in their appendix,
$\epsilon_e$ and $\epsilon_B$ are the equipartition factors for the energy in electrons and
magnetic field in the shock, respectively.
For the photons with frequency higher than $\hat{\nu}^{\prime}$, the Compton parameter
should be suppressed significantly since it is the Klein–Nishina
regime,
where $\hat{\nu}^{\prime}$ is governed by
$\gamma_e^{\prime} h \hat{\nu}^{\prime} \sim \Gamma m_e c^2$.
Since we have consider the case of $p>2$, for the slow-cooling case ($\gamma_m<\gamma_c$),
\begin{equation}
\eta_{\rm KN} \sim
\left\{
\renewcommand{\arraystretch}{1.5}
\begin{array}{lr}
0, &  \hat{\gamma}^{\prime}_e<\gamma^{\prime}_m, \\
\dfrac{\hat{\gamma}_e^{\prime 3-p}-\gamma_m^{\prime 3-p}}{[1/(p-2)]\gamma_c^{3-p}-\gamma_m^{\prime 3-p}},
& \gamma^{\prime}_m<\hat{\gamma}^{\prime}_e <\gamma^{\prime}_c, \\
1-\dfrac{(3-p)\gamma^{\prime}_c \hat{\gamma}_e^{\prime 2-p}}{\gamma_c^{3-p}-(p-2) \gamma_m^{\prime 3-p}},
& \gamma^{\prime}_c<\hat{\gamma}^{\prime}_e,
\end{array}
\right.
\end{equation}
also, for the fast-cooling case ($\gamma^{\prime}_m>\gamma^{\prime}_c$),
\begin{equation}
\eta_{\rm KN} \sim
\left\{
\renewcommand{\arraystretch}{1.5}
\begin{array}{lr}
0,
& \hat{\gamma}^{\prime}_e <\gamma^{\prime}_c, \\
\dfrac{\hat{\gamma}^{\prime}_e-\gamma^{\prime}_c}{[(p-1)/(p-2)]\gamma^{\prime}_m-\gamma^{\prime}_c},
& \gamma^{\prime}_c<\hat{\gamma}^{\prime}_e<\gamma^{\prime}_m, \\
1-\dfrac{\gamma_m^{\prime p-1}\hat{\gamma}_e^{\prime 2-p}}{(p-1)\gamma^{\prime}_m-(p-2) \gamma^{\prime}_c},
& \gamma^{\prime}_m<\hat{\gamma}^{\prime}_e,
\end{array}
\right.
\end{equation}
where $\hat{\gamma}^{\prime}_e=\sqrt{4\pi m_e c\hat{\nu}^{\prime}/3q_e B^{\prime}}$,
and $\gamma^{\prime}_c=6 \pi m_e c/(\sigma_{\rm T}\Gamma {B'}^2 t')$
is the efficient cooling Lorentz factor of electrons with
$\sigma_{\rm T}$ being the Thomson scattering cross section.

The electron populations in both FS and RS has treated using the method described above.
For electrons in RS,
$n(r)$ in Equation~{\ref{eq:Qinj}} should be considered as $n(r)=n_4$.
Unlike the continuous injection of electrons in the case of FS,
no new electrons are injected after the crossing time
$t_{\times}^{\prime}$ of RS which has traversed all the matter in the jet shell,
completely replacing zone 4 with zone 3.

Then, the magnetic field behind the FS
is $B^{\prime}={[32\pi { \epsilon_{B,f}}{n(r)}]^{1/2}}\Gamma c$,
but has computed as $B^{\prime}=\sqrt{8\pi \epsilon_{B,r} e_3}$
for the RS, where $e_3=e_2=4\gamma_2^2 n(r)m_p c^2$
when $t'<t_{\times}^{\prime}$,
and $e_3\propto r^2_{\times}\Delta_{\times}/(r^2\Delta)=(r_{\times}/r)^3\gamma_2/\gamma_{\times}$
when $t'>t_{\times}^{\prime}$, respectively.

Additionally, one can have
$\epsilon=\epsilon_{\rm rad}\epsilon_e$
with $\epsilon_{\rm rad}=\min \{1,(\gamma^{\prime}_m/\gamma^{\prime}_c)^{(p-2)}\}$
(\citealp{Sari_2001_Esin_apj_v548.p787..799,Fan_2008_Piran_mnras_v384.p1483..1501}),
for Equations~\ref{eq:RS}, \ref{eq:diff_U} and \ref{eq:Y}.

\subsection{Radiation Mechanism}
In the X-ray/optical/radio bands,
the main radiation mechanism of the electrons in GRB jets
is synchrotron radiation
(\citealp{Sari_1998_Piran_apj_v497.p17..20L,Sari-1999-Piran-ApJ...517L.109S}).
The emitted spectral power of synchrotron radiation
at a given frequency $\nu'$ of a single electron is
\begin{equation}
P^{\prime}_e(\nu^{\prime}, \gamma^{\prime}_e)
=\frac{\sqrt{3} q_e^3 B^{\prime}}{m_e c^2}
F\left(\frac{\nu'}{\nu'_c}\right),
\end{equation}
where we have assumed that the direction of the magnetic field is perpendicular to the electron velocity.
$F(x)=x\int_{x}^{+\infty}K_{5/3}(k)dk$,
$K_{5/3}(k)$ is the modified Bessel function of $5/3$ order,
and $\nu'_{\rm c}=3q_e B'{\gamma'_e}^2/(4\pi m_e c)$, respectively.
Thus, the spectral power of synchrotron radiation of electrons
${dN_e}/d\gamma_e^{\prime}$ at a given frequency $\nu'$ is
\begin{equation}
P'_{\rm syn}(\nu')=
\int_{\gamma^{\prime}_m}^{\gamma^{\prime}_M}
P'_e(\nu')\frac{dN_e}{d\gamma_e^{\prime}}d\gamma^{\prime}_e,
\end{equation}

The photon seed spectra of synchrotron radiation is then be calculated as
\begin{equation}\label{seed_nu}
n'_{\gamma,\rm syn}\left(\nu^{\prime}\right) \simeq
\frac{T^{\prime}}{h \nu^{\prime}} \int_{\gamma_m^{\prime}}^{\gamma_M^{\prime}}
P'_e(\nu')
\frac{dN_e / d\gamma_e^{\prime}}{4\pi r^2 \Delta}
d\gamma_e^{\prime}
=\frac{P'_{\rm syn}(\nu')}{4\pi r^2 c h\nu'}
\end{equation}
where $T^{\prime}\approx\Delta/c$
is the dynamical timescale of synchrotron radiation photons,
$\dfrac{dN_e / d\gamma_e^{\prime}}{4\pi r^2 \Delta}$
is the co-moving electron number density,
and $\Delta\approx r/12\Gamma$ is the co-moving width of the jet shell.

The emission of the SSC process is calculated based on the electron spectrum
${dN_e}/d\gamma_e^{\prime}$ and target seed photons of synchrotron radiation
$n_{\gamma,\rm syn}^{\prime}\left(\nu_t^{\prime}\right)$
from the synchrotron radiation
\citep[e.g.,][]{Fan_2008_Piran_mnras_v384.p1483..1501,Nakar_2009_Ando_apj_v703.p675..691,
Geng_2018_Huang_apjs_v234.p3..3}
\begin{equation}
P_{\rm SSC}^{\prime}\left(\nu^{\prime}\right)
=\frac{3 \sigma_{\rm T}ch\nu^{\prime}}{4}
\int_{\nu_{\min}^{\prime}}^{\nu_{\max}^{\prime}}
\frac{n_{\gamma, \rm syn}^{\prime}\left(\nu_t^{\prime}\right) d\nu_t^{\prime}}{\nu_t^{\prime}}
\int_{\gamma_m^{\prime}}^{\gamma_M^{\prime}} \frac{F(q, g)}{\gamma_e^{\prime 2}}
\frac{dN_e}{d\gamma_e^{\prime}} d\gamma_e^{\prime},
\end{equation}
where
$F(q, g)=2q\ln q+(1+2q)(1-q)+8q^2 g^2(1-q)(1+4qg)$,
$q=w/4g(1-w)$,
$g=\gamma_e^{\prime}h\nu_t^{\prime}/m_e c^2$,
and $w=h\nu^{\prime}/\gamma_e^{\prime} m_e c^2$, respectively.
One can also derive the photon seed spectra of SSC radiation
\begin{equation}
n'_{\gamma,\rm SSC}\left(\nu^{\prime}\right)=
\frac{P'_{\rm SSC}(\nu')}{4\pi r^2 c h\nu'}.
\end{equation}
The total spectral power and photon seed spectra then be
$P'_{\rm tot}(\nu')=P'_{\rm syn}(\nu')+P'_{\rm SSC}(\nu')$
and
$n'_{\gamma,\rm tot}=n'_{\gamma,\rm syn}+n'_{\gamma,\rm SSC}$,
respectively.

We also considered the $\gamma\gamma$ annihilation effects
(e.g., \citealp{Gould_1967_Schreder_PhRv..155.1404,Geng_2018_Huang_apjs_v234.p3..3,
Huang_2022__apj_v931.p150..150}).
Since the cross section of $\gamma\gamma$ annihilation reads
\begin{equation}
\sigma_{\gamma\gamma}=\frac{3}{16} \sigma_{T}\left(1-\beta_{\rm cm}^2\right)
\left[2 \beta_{\rm cm}\left(\beta_{\rm cm}^2-2\right)+\left(3-\beta_{\rm cm}^4\right)
\ln \frac{1+\beta_{\rm cm}}{1-\beta_{\rm cm}}\right],
\end{equation}
and
\begin{equation}
\beta_{\rm cm}=\frac{v}{c}=\sqrt{1-\frac{2\left(m_e c^2\right)^2}{h\nu^{\prime} h\nu^{\prime}_{t}(1-\mu)}},
\end{equation}
where $\beta_{\rm cm}$ is the dimensionless velocity of electron (positron) in the center-of-momentum frame,
$\mu=\cos{\theta}$ with $\theta$ being the angle of collision photons, and
$\nu^{\prime}_{t}$ is the frequency of target photons from synchrotron and IC radiation.
It is obvious that there is a physical value only when
$h\nu^{\prime} h\nu^{\prime}_t(1-\mu) \geqslant 2\left(m_e c^2\right)^2$.
Thus, the optical depth for a gamma-ray photon with frequency $\nu^{\prime}$
is expressed as \citep[e.g.,][]{Murase_2011_Toma_apj_v732.p77..77}
\begin{equation}
\tau^{\gamma\gamma}(\nu^{\prime})=\frac{\Delta}{2} \int_{-1}^{1} (1-\mu)d\mu
\int\sigma_{\gamma\gamma} n'_{\gamma,\rm tot}(\nu^{\prime}_t) d\nu^{\prime}_t.
\end{equation}

The synchrotron self-absorption (SSA) effect is also considered
in our numerical calculations. The optical depth is the
function of both the electron distribution
and synchrotron radiation power of single electron,
\begin{equation}
\tau^{\rm SSA}(\nu^{\prime})=-\frac{1}{8 \pi m_e \nu^{\prime 2}}
\int \gamma^{\prime 2}_e P^{\prime}_e(\nu^{\prime}, \gamma^{\prime}_e)
\frac{\partial}{\partial \gamma^{\prime}_e}
\left[\frac{dN_e/d\gamma_e^{\prime}}{\gamma_e^{\prime 2}}
\right] d\gamma_e^{\prime}.
\end{equation}

Finally, the intrinsic spectral power of afterglow is
\begin{equation}
P^{\prime \rm in}_{\rm tot}(\nu^{\prime})=
P^{\prime}_{\rm tot}(\nu^{\prime})
\frac{1-e^{-\left[\tau^{\gamma\gamma}(\nu^{\prime})
+\tau^{\rm SSA}(\nu^{\prime})\right]}}
{\tau^{\gamma\gamma}(\nu^{\prime})
+\tau^{\rm SSA}(\nu^{\prime})},
\end{equation}
note that $\gamma\gamma$ annihilation and SSA
dominate the absorption of
high-energy and low-energy photons, respectively.

\subsection{The Geometric and Observational Effects}
GRBs are believed to be generated by
the motion of ultra-relativistic jets within a half-opening angle $\theta_j$.
Due to geometric and relativistic Doppler effects,
the observed radiation spectrum cannot be
simply reproduced by the intrinsic spectrum, i.e.,
$P^{\prime \rm in}_{\rm tot}(\nu^{\prime})$.
In this work, we set the GRB jet as an on-axis-observed jet, i.e.,
$\theta_v=0$.
Assuming that the observed frequency is $\nu_{\rm obs}$, the frequency transformed to the comoving frame is denoted as
$\nu^{\prime}(\nu_{\rm obs})=(1+z)\nu_{\rm obs}/\mathcal{D}$, where $z$ is the redshift,
$\mathcal{D}=\dfrac{1}{\Gamma(1-\beta\cos\Theta)}$
is the Doppler factor,
and $\Theta$ represents the angle between the direction of motion
of a fluid element in the jet and the line-of-sight.

If we take the equal-arrival-time surface (EATS)
effect into account \citep{Waxman_1997_ApJ...485L...5W}, the observed intrinsic spectral flux should be
\begin{equation}
F^{\rm in}_{\rm tot}\left(\nu_{\rm obs}\right)=
\frac{1+z}{4 \pi D_{L}^2} \int_{0}^{\theta_j}
P^{\prime \rm in}_{\rm tot}\left[\nu^{\prime}\left(\nu_{\rm obs}\right)\right]
\mathcal{D}^3 \frac{\sin \theta}{2} d\theta,
\end{equation}
where $D_L$ is the luminosity distance from burst to the Earth.
The integration of $\theta$ is performed over an elliptical surface
that photons emitted from the surface have the same arrival time for an observer
\citep{Geng_2018_Huang_apjs_v234.p3..3},
\begin{equation}
t_{\mathrm{obs}}=
(1+z) \int \frac{1-\beta\cos\Theta}{\beta c} dr \equiv \rm{const}.
\end{equation}

Gamma-ray photons traveling through the universe
may undergo $\gamma\gamma$ annihilation with extragalactic background light (EBL) photons,
resulting in additional absorption of high-energy photons.
When considering the EBL absorption,
the finally observed flux reads
\begin{equation}
F^{\rm obs}_{\rm tot}\left(\nu_{\rm obs}\right)=
F^{\rm in}_{\rm tot}\left(\nu_{\rm obs}\right)\exp
\left[-\tau^{\rm EBL}(\nu_{\rm obs},z)\right].
\end{equation}

\subsection{Numerical Method}
The evolution of dynamics is a system of ordinary differential equations
which is solved using the fourth-order Runge-Kutta method.
For the distribution equation of electrons, which is a partial differential equation,
a fully implicit first order scheme is implemented
\citep{Chang_1970_Cooper_JournalofComputationalPhysics_v6.p1..16,
Chiaberge_1999_Ghisellini_mnras_v306.p551..560,Huang_2022__apj_v931.p150..150},
with grid transformations to improve computational accuracy
\citep[e.g.,][]{Geng_2018_Huang_apjs_v234.p3..3}.
During computations related to the radiation spectrum,
the Euler method and linear interpolation are used.
Therefore, the computed results used for fitting are believed to have first-order accuracy.

In this study, we assume that the jet is structured.
This implies that the development of the jet at different angles is not connected.
Consequently, we calculate the intrinsic radiation spectrum for each angle independently,
and the final result is obtained by using the interpolation technique.
We do not take into account any potential lateral expansion effects.
This simplification is suitable when the jet is highly relativistic;
however, it may lead to an overestimation of the flux after the jet break time.

\acknowledgments
The Rainbow Bridge of Asgard in Norse mythology,
has inspired R.J. to name the {\tt ASGARD} package.
It traverses the universe with shining, just like the GRB afterglow.
This work was supported by
the National Natural Science Foundation of China (grant No. 11833003).
This work used data and software provided by the Fermi Science Support Center
and data supplied by the UK Swift Science Data Centre at the University of Leicester.


\software{\texttt{Matplotlib}
\citep{Hunter_2007__ComputinginScienceandEngineering_v9.p90..95},
\texttt{NumPy}
\citep{Harris_2020_Millman_Nature_v585.p357..362},
\texttt{emcee}
\citep{ForemanMackey_2013_Hogg_pasp_v125.p306..312},
\texttt{corner}
\citep{ForemanMackey_2016__TheJournalofOpenSourceSoftware_v1.p24..24},
\texttt{Astropy}
\citep{Collaboration_2013_Robitaille_aap_v558.p33..33A},
\texttt{extinction}
\citep{Barbary_2016___v.p..},
\texttt{SciPy}
\citep{Virtanen_2020_Gommers_NatureMethods_v17.p261..272},
\texttt{GetDist}
\citep{Lewis_2019___v.p..}}
\clearpage


\begin{thebibliography}{}
\expandafter\ifx\csname natexlab\endcsname\relax\def\natexlab#1{#1}\fi
\providecommand{\url}[1]{\href{#1}{#1}}
\providecommand{\dodoi}[1]{doi:~\href{http://doi.org/#1}{\nolinkurl{#1}}}
\providecommand{\doeprint}[1]{\href{http://ascl.net/#1}{\nolinkurl{http://ascl.net/#1}}}
\providecommand{\doarXiv}[1]{\href{https://arxiv.org/abs/#1}{\nolinkurl{https://arxiv.org/abs/#1}}}

\bibitem[{Abdalla {et~al.}(2019)Abdalla, Adam, Aharonian,
  {\.Z}ywucka, de~Palma, Axelsson, \&
  Roberts}]{Abdalla_2019_Adam_nat_v575.p464..467}
Abdalla, H., Adam, R., Aharonian, F., {et~al.} 2019, \nat, 575, 464,
  \dodoi{10.1038/s41586-019-1743-9}

\bibitem[Abe et al.(2023)]{Abe_2023_Abe_mnras.tmp.3105A}
Abe, H., Abe, S., Acciari, V.~A., et al.\ 2023, \mnras.
\dodoi{10.1093/mnras/stad2958}


\bibitem[{Acciari {et~al.}(2011)Acciari, Aliu, Arlen, Williams, \&
  Wood}]{Acciari_2011_Aliu_apj_v743.p62..62}
Acciari, V.~A., Aliu, E., Arlen, T., {et~al.} 2011, \apj, 743, 62,
  \dodoi{10.1088/0004-637X/743/1/62}

\bibitem[{Alexander {et~al.}(2017)Alexander, Laskar, Berger, Guidorzi,
  Dichiara, Fong, Gomboc, Kobayashi, Kopac, Mundell, Tanvir, \&
  Williams}]{Alexander_2017_Laskar_apj_v848.p69..69}
Alexander, K.~D., Laskar, T., Berger, E., {et~al.} 2017, \apj, 848, 69,
  \dodoi{10.3847/1538-4357/aa8a76}

\bibitem[{Alfaro {et~al.}(2017)Alfaro, Alvarez, {\'A}lvarez, Arceo,
 Zhou, \& Collaboration}]{Alfaro_2017_Alvarez_apj_v843.p88..88}
Alfaro, R., Alvarez, C., {\'A}lvarez, J.~D., {et~al.} 2017, \apj, 843, 88,
  \dodoi{10.3847/1538-4357/aa756f}

\bibitem[{Aliu {et~al.}(2014)Aliu, Aune, Barnacka, Veres, \& Zhu}]
{Aliu_2014_Aune_apjl_v795.p3..3L}
Aliu, E., Aune, T., Barnacka, A., {et~al.} 2014, \apjl, 795, L3,
  \dodoi{10.1088/2041-8205/795/1/L3}

\bibitem[{An {et~al.}(2023)An, Antier, Bi, Zhou, Zhou, \&
  Zhu}]{An_2023_arXiv230301203A}
An, Z.-H., Antier, S., Bi, X.-Z., {et~al.} 2023, arXiv e-prints,
  arXiv:2303.01203, \dodoi{10.48550/arXiv.2303.01203}

\bibitem[{{Astropy Collaboration} {et~al.}(2013){Astropy Collaboration},
  Robitaille, Tollerud, Greenfield, Droettboom, Bray, Aldcroft, Davis,
  Ginsburg, Price-Whelan, Kerzendorf, Conley, Crighton, Barbary, Muna,
  Ferguson, Grollier, Parikh, Nair, Unther, Deil, Woillez, Conseil, Kramer,
  Turner, Singer, Fox, Weaver, Zabalza, Edwards, Azalee~Bostroem, Burke, Casey,
  Crawford, Dencheva, Ely, Jenness, Labrie, Lim, Pierfederici, Pontzen, Ptak,
  Refsdal, Servillat, \&
  Streicher}]{Collaboration_2013_Robitaille_aap_v558.p33..33A}
{Astropy Collaboration}, Robitaille, T.~P., Tollerud, E.~J., {et~al.} 2013,
  \aap, 558, A33, \dodoi{10.1051/0004-6361/201322068}

\bibitem[{Barbary(2016)}]{Barbary_2016___v.p..}
Barbary, K. 2016, extinction v0.3.0,  Zenodo, \dodoi{10.5281/zenodo.804967}

\bibitem[{Barniol~Duran(2014)}]{BarniolDuran_2014__mnras_v442.p3147..3154}
Barniol~Duran, R. 2014, \mnras, 442, 3147, \dodoi{10.1093/mnras/stu1070}

\bibitem[Beniamini \& van der Horst(2017)]{Beniamini_2017_vanderHorst_mnras_v472.p3161B}
Beniamini, P. \& van der Horst, A.~J.\ 2017, \mnras, 472, 3161. \dodoi{10.1093/mnras/stx2203}

\bibitem[{Berti \& Carosi(2022)}]{Berti_2022_Carosi_Galaxies_v10.p67..67}
Berti, A., \& Carosi, A. 2022, Galaxies, 10, 67,
  \dodoi{10.3390/galaxies10030067}

\bibitem[{Berti \& Group(2017)}]{Berti_2017_Group__v324.p70..73}
Berti, A., \& Group, M.~G. 2017, in New Frontiers in Black Hole Astrophysics,
  ed. A.~{Gomboc}, Vol. 324, 70--73, \dodoi{10.1017/S1743921317001442}

\bibitem[Blanchard et al.(2023)]{Blanchard_2023_Villar_arXiv230814197B}
Blanchard, P.~K., Villar, V.~A., Chornock, R., et al.\ 2023,
\newblock \doarXiv{2308.06994}

\bibitem[{Blandford \&
  McKee(1976)}]{Blandford_1976_McKee_PhysicsofFluids_v19.p1130..1138}
Blandford, R.~D., \& McKee, C.~F. 1976, Physics of Fluids, 19, 1130,
  \dodoi{10.1063/1.861619}

\bibitem[{Bright {et~al.}(2023)Bright, Rhodes, Farah, Fender, van~der Horst,
  Leung, Williams, Anderson, Atri, DeBoer, Giarratana, Green, Heywood, Lenc,
  Murphy, Pollak, Premnath, Scott, Sheikh, Siemion, \&
  Titterington}]{Bright_2023_Rhodes_NatureAstronomy_v.p..}
Bright, J.~S., Rhodes, L., Farah, W., {et~al.} 2023, Nature Astronomy,
  \dodoi{10.1038/s41550-023-01997-9}

\bibitem[{Cantiello {et~al.}(2009)Cantiello, Langer, Brott, de~Koter, Shore,
  Vink, Voegler, Lennon, \& Yoon}]{Cantiello_2009_Langer_aap_v499.p279..290}
Cantiello, M., Langer, N., Brott, I., {et~al.} 2009, \aap, 499, 279,
  \dodoi{10.1051/0004-6361/200911643}

\bibitem[Cao et al.(2023)]{Cao_2023_Aharonian_SciA....9J2778C} Cao, Z., Aharonian, F., An, Q., et al.\ 2023, Science Advances, 9, eadj2778. doi:10.1126/sciadv.adj2778


\bibitem[{Chang \&
  Cooper(1970)}]{Chang_1970_Cooper_JournalofComputationalPhysics_v6.p1..16}
Chang, J.~S., \& Cooper, G. 1970, Journal of Computational Physics, 6, 1,
  \dodoi{10.1016/0021-9991(70)90001-X}

\bibitem[{Chen \& Liu(2021)}]{Chen_2021_Liu_mnras_v504.p1759..1771}
Chen, Q., \& Liu, X.~W. 2021, \mnras, 504, 1759, \dodoi{10.1093/mnras/stab946}

\bibitem[{Chevalier {et~al.}(2004)Chevalier, Li, \&
  Fransson}]{Chevalier_2004_Li_apj_v606.p369..380}
Chevalier, R.~A., Li, Z.-Y., \& Fransson, C. 2004, \apj, 606, 369,
  \dodoi{10.1086/382867}

\bibitem[{Chiaberge \&
  Ghisellini(1999)}]{Chiaberge_1999_Ghisellini_mnras_v306.p551..560}
Chiaberge, M., \& Ghisellini, G. 1999, \mnras, 306, 551,
  \dodoi{10.1046/j.1365-8711.1999.02538.x}

\bibitem[{Corsi {et~al.}(2014)Corsi, Ofek, Gal-Yam, Frail, Kulkarni, Fox,
  Kasliwal, Sullivan, Horesh, Carpenter, Maguire, Arcavi, Cenko, Cao, Mooley,
  Pan, Sesar, Sternberg, Xu, Bersier, James, Bloom, \&
  Nugent}]{Corsi_2014_Ofek_apj_v782.p42..42}
Corsi, A., Ofek, E.~O., Gal-Yam, A., {et~al.} 2014, \apj, 782, 42,
  \dodoi{10.1088/0004-637X/782/1/42}

\bibitem[Cunningham et al.(2020)]{Cunningham_2020_Cenko_apj_v904.p166..166}
Cunningham, V., Cenko, S.~B., Ryan, G., et al.\ 2020, \apj, 904, 166.
\dodoi{10.3847/1538-4357/abc2cd}

\bibitem[{Dai \& Lu(2002)}]{Dai_2002_Lu_apjl_v565.p87..90L}
Dai, Z.~G., \& Lu, T. 2002, \apjl, 565, L87, \dodoi{10.1086/339418}

\bibitem[{Dai \& Wu(2003)}]{Dai_2003_Wu_apjl_v591.p21..24L}
Dai, Z.~G., \& Wu, X.~F. 2003, \apjl, 591, L21, \dodoi{10.1086/377037}

\bibitem[Duncan et al.(2023)]{Duncan_2023_vanderHorst_mnras_v518.p1522D}
Duncan, R.~A., van der Horst, A.~J., \& Beniamini, P.\ 2023, \mnras, 518, 1522.
\dodoi{10.1093/mnras/stac3172}

\bibitem[{Eichler \& Waxman(2005)}]{Eichler_2005_Waxman_apj_v627.p861..867}
Eichler, D., \& Waxman, E. 2005, \apj, 627, 861, \dodoi{10.1086/430596}

\bibitem[{Eldridge {et~al.}(2006)Eldridge, Genet, Daigne, \&
  Mochkovitch}]{Eldridge_2006_Genet_mnras_v367.p186..200}
Eldridge, J.~J., Genet, F., Daigne, F., \& Mochkovitch, R. 2006, \mnras, 367,
  186, \dodoi{10.1111/j.1365-2966.2005.09938.x}

\bibitem[{Fan \& Piran(2006)}]{Fan_2006_Piran_mnras_v369.p197..206}
Fan, Y.~Z., \& Piran, T. 2006, \mnras, 369, 197,
  \dodoi{10.1111/j.1365-2966.2006.10280.x}

\bibitem[{Fan {et~al.}(2008)Fan, Piran, Narayan, \&
  Wei}]{Fan_2008_Piran_mnras_v384.p1483..1501}
Fan, Y.~Z., Piran, T., Narayan, R., \& Wei, D.-M. 2008, \mnras, 384, 1483,
  \dodoi{10.1111/j.1365-2966.2007.12765.x}

\bibitem[{Filgas {et~al.}(2011)Filgas, Kr{\"u}hler, Greiner, Rau, Palazzi,
  Klose, Schady, Rossi, Afonso, Antonelli, Clemens, Covino, D'Avanzo,
  K{\"u}pc{\"u}~Yolda{\c{s}}, Nardini, Nicuesa~Guelbenzu, Olivares, Updike, \&
  Yolda{\c{s}}}]{Filgas_2011_Kruehler_aap_v526.p113..113A}
Filgas, R., Kr{\"u}hler, T., Greiner, J., {et~al.} 2011, \aap, 526, A113,
  \dodoi{10.1051/0004-6361/201015320}

\bibitem[{Foreman-Mackey(2016)}]{ForemanMackey_2016__TheJournalofOpenSourceSoftware_v1.p24..24}
Foreman-Mackey, D. 2016, The Journal of Open Source Software, 1, 24,
  \dodoi{10.21105/joss.00024}

\bibitem[{Foreman-Mackey {et~al.}(2013)Foreman-Mackey, Hogg, Lang, \&
  Goodman}]{ForemanMackey_2013_Hogg_pasp_v125.p306..312}
Foreman-Mackey, D., Hogg, D.~W., Lang, D., \& Goodman, J. 2013, \pasp, 125,
  306, \dodoi{10.1086/670067}

\bibitem[{Fukami {et~al.}(2021)Fukami, Berti, Loporchio, Will, Wunderlich, Yamamoto, \&
  Zarić}]{Fukami-2022-Berti-icrc.confE.788F}
Fukami, S., Berti, A., Loporchio, S., {et~al.} 2021, in Proceedings of 37th
  International Cosmic Ray Conference {\textemdash} PoS(ICRC2021), Vol. 395,
  788, \dodoi{10.22323/1.395.0788}

\bibitem[{Fukami {et~al.}(2023)Fukami, Abe, Abe, Abhir, Will, Wunderlich, \& and}]
{Fukami_2023_Abe__v.p..}
Fukami, S., Abe, H., Abe, S., {et~al.} 2023, in Proceedings of 38th
  International Cosmic Ray Conference {\textemdash} {PoS}({ICRC}2023) (Sissa
  Medialab), \dodoi{10.22323/1.444.0791}

\bibitem[{Fukushima {et~al.}(2017)Fukushima, To, Asano, \&
  Fujita}]{Fukushima_2017_To_apj_v844.p92..92}
Fukushima, T., To, S., Asano, K., \& Fujita, Y. 2017, \apj, 844, 92,
  \dodoi{10.3847/1538-4357/aa7b83}

\bibitem[{Fulton {et~al.}(2023)Fulton, Smartt, Rhodes, Huber, Villar, Moore,
  Srivastav, Schultz, Chambers, Izzo, Hjorth, Chen, Nicholl, Foley, Rest,
  Smith, Young, Sim, Bright, Zenati, de~Boer, Bulger, Fairlamb, Gao, Lin, Lowe,
  Magnier, Smith, Wainscoat, Coulter, Jones, Kilpatrick, McGill, Ramirez-Ruiz,
  Lee, Narayan, Ramakrishnan, Ridden-Harper, Singh, Wang, Kong, Ngeow, Pan,
  Yang, Davis, Piro, Rojas-Bravo, Sommer, \&
  Yadavalli}]{Fulton_2023_Smartt_apjl_v946.p22..22L}
Fulton, M.~D., Smartt, S.~J., Rhodes, L., {et~al.} 2023, \apjl, 946, L22,
  \dodoi{10.3847/2041-8213/acc101}

\bibitem[{Gao {et~al.}(2015)Gao, Wang, M{\'e}sz{\'a}ros, \&
  Zhang}]{Gao_2015_Wang_apj_v810.p160..160}
Gao, H., Wang, X.-G., M{\'e}sz{\'a}ros, P., \& Zhang, B. 2015, \apj, 810, 160,
  \dodoi{10.1088/0004-637X/810/2/160}

\bibitem[{Geng {et~al.}(2018)Geng, Huang, Wu, Zhang, \&
  Zong}]{Geng_2018_Huang_apjs_v234.p3..3}
Geng, J.-J., Huang, Y.-F., Wu, X.-F., Zhang, B., \& Zong, H.-S. 2018, \apjs,
  234, 3, \dodoi{10.3847/1538-4365/aa9e84}

\bibitem[{Geng {et~al.}(2019)Geng, Zhang, K{\"o}lligan, Kuiper, \&
  Huang}]{Geng_2019_Zhang_apjl_v877.p40..40L}
Geng, J.-J., Zhang, B., K{\"o}lligan, A., Kuiper, R., \& Huang, Y.-F. 2019,
  \apjl, 877, L40, \dodoi{10.3847/2041-8213/ab224b}

\bibitem[{Ghirlanda {et~al.}(2019)Ghirlanda, Salafia, Paragi, Venturi, Vergani, \&
  Zhang}]{Ghirlanda_2019_Salafia_Science_v363.p968..971}
Ghirlanda, G., Salafia, O.~S., Paragi, Z., {et~al.} 2019, Science, 363, 968,
  \dodoi{10.1126/science.aau8815}

\bibitem[{Ghisellini {et~al.}(2007)Ghisellini, Ghirlanda, Nava, \&
  Firmani}]{Ghisellini_2007_Ghirlanda_apjl_v658.p75..78L}
Ghisellini, G., Ghirlanda, G., Nava, L., \& Firmani, C. 2007, \apjl, 658, L75,
  \dodoi{10.1086/515570}

\bibitem[{Gill \& Granot(2022)}]{Gill_2022_Granot_Galaxies_v10.p74..74}
Gill, R., \& Granot, J. 2022, Galaxies, 10, 74,
  \dodoi{10.3390/galaxies10030074}

\bibitem[{Gill \& Granot(2023)}]{Gill_2023_Granot_mnras_v524.p78..83L}
---. 2023, \mnras, 524, L78, \dodoi{10.1093/mnrasl/slad075}

\bibitem[{Gottlieb {et~al.}(2022)Gottlieb, Moseley, Ramirez-Aguilar,
  Murguia-Berthier, Liska, \&
  Tchekhovskoy}]{Gottlieb_2022_Moseley_apjl_v933.p2..2L}
Gottlieb, O., Moseley, S., Ramirez-Aguilar, T., {et~al.} 2022, \apjl, 933, L2,
  \dodoi{10.3847/2041-8213/ac7728}

\bibitem[{Gottlieb {et~al.}(2021)Gottlieb, Nakar, \&
  Bromberg}]{Gottlieb_2021_Nakar_mnras_v500.p3511..3526}
Gottlieb, O., Nakar, E., \& Bromberg, O. 2021, \mnras, 500, 3511,
  \dodoi{10.1093/mnras/staa3501}

\bibitem[{{Gould} \& {Schr{\'e}der}(1967)}]{Gould_1967_Schreder_PhRv..155.1404}
{Gould}, R.~J., \& {Schr{\'e}der}, G.~P. 1967, Physical Review, 155, 1408,
  \dodoi{10.1103/PhysRev.155.1408}

\bibitem[{{H.~E.~S.~S. Collaboration} {et~al.}(2021){H.~E.~S.~S.
  Collaboration}, {Abdalla}, {Aharonian}, {Ait Benkhali}, {Ang{\"u}ner},
  {Arcaro}, {Armand}, {Armstrong}, {Ashkar}, {Backes}, \&
  et~al.}]{Collaboration_2021_Abdalla_Science_v372.p1081..1085}
{H.~E.~S.~S. Collaboration}, {Abdalla}, H., {Aharonian}, F., {et~al.} 2021,
  Science, 372, 1081, \dodoi{10.1126/science.abe8560}

\bibitem[{Harris {et~al.}(2020)Harris, Millman, van~der Walt, Gommers,
  Virtanen, Cournapeau, Wieser, Taylor, Berg, Smith, Kern, Picus, Hoyer, van
  Kerkwijk, Brett, Haldane, del R{\'{\i}}o, Wiebe, Peterson,
  G{\'{e}}rard-Marchant, Sheppard, Reddy, Weckesser, Abbasi, Gohlke, \&
  Oliphant}]{Harris_2020_Millman_Nature_v585.p357..362}
Harris, C.~R., Millman, K.~J., van~der Walt, S.~J., {et~al.} 2020, Nature, 585,
  357, \dodoi{10.1038/s41586-020-2649-2}

\bibitem[{Hoischen {et~al.}(2017)Hoischen, Balzer, Bissaldi, F{\"u}{\ss}ling,
  Garrigoux, Gottschall, Holler, Mitchell, O'Brien, Parsons, P{\"u}hlhofer,
  Rowell, Sch{\"u}ssler, Tam, Wagner, \&
  Collaboration}]{Hoischen_2017_Balzer__v301.p636..636}
Hoischen, C., Balzer, A., Bissaldi, E., {et~al.} 2017, in International Cosmic
  Ray Conference, Vol. 301, 35th International Cosmic Ray Conference
  (ICRC2017), 636, \dodoi{10.22323/1.301.0636}

\bibitem[{Hotokezaka {et~al.}(2019)Hotokezaka, Nakar, Gottlieb, Nissanke,
  Masuda, Hallinan, Mooley, \&
  Deller}]{Hotokezaka_2019_Nakar_NatureAstronomy_v3.p940..944}
Hotokezaka, K., Nakar, E., Gottlieb, O., {et~al.} 2019, Nature Astronomy, 3,
  940, \dodoi{10.1038/s41550-019-0820-1}

\bibitem[{Huang {et~al.}(2023)Huang, Huang, Cheng, Ren, Zhang, \&
  Liang}]{Huang_2023_Huang_apj_v947.p84..84}
Huang, J.-K., Huang, X.-L., Cheng, J.-G., {et~al.} 2023, \apj, 947, 84,
  \dodoi{10.3847/1538-4357/acc85f}

\bibitem[{{Huang}(2022)}]{Huang_2022__apj_v931.p150..150}
{Huang}, Y. 2022, \apj, 931, 150, \dodoi{10.3847/1538-4357/ac6d52}

\bibitem[{{Huang} {et~al.}(2022){Huang}, {Hu}, {Chen}, \& et~al}]{GCN32677}
{Huang}, Y., {Hu}, S.-C., {Chen}, S.-Z., \& et~al. 2022, GRB Coordinates
  Network, 32677, 1

\bibitem[{Huang {et~al.}(2004)Huang, Wu, Dai, Ma, \&
  Lu}]{Huang_2004_Wu_apj_v605.p300..306}
Huang, Y.~F., Wu, X.~F., Dai, Z.~G., Ma, H.~T., \& Lu, T. 2004, \apj, 605, 300,
  \dodoi{10.1086/382202}

\bibitem[{{Hunter}(2007)}]{Hunter_2007__ComputinginScienceandEngineering_v9.p90..95}
{Hunter}, J.~D. 2007, Computing in Science and Engineering, 9, 90,
  \dodoi{10.1109/MCSE.2007.55}

\bibitem[{Isravel {et~al.}(2023)Isravel, Begue, \&
  Pe'er}]{Isravel_2023_Begue__v.p..}
Isravel, H., Begue, D., \& Pe'er, A. 2023.
\newblock \doarXiv{2308.06994}

\bibitem[{Japelj {et~al.}(2014)Japelj, Kopa{\v{c}}, Kobayashi, Harrison,
  Guidorzi, Virgili, Mundell, Melandri, \&
  Gomboc}]{Japelj_2014_Kopac_apj_v785.p84..84}
Japelj, J., Kopa{\v{c}}, D., Kobayashi, S., {et~al.} 2014, \apj, 785, 84,
  \dodoi{10.1088/0004-637X/785/2/84}

\bibitem[{Kangas {et~al.}(2020)Kangas, Fruchter, Cenko, Corsi,
  de~Ugarte~Postigo, Pe'er, Vogel, Cucchiara, Gompertz, Graham, Levan, Misra,
  Perley, Racusin, \& Tanvir}]{Kangas_2020_Fruchter_apj_v894.p43..43}
Kangas, T., Fruchter, A.~S., Cenko, S.~B., {et~al.} 2020, \apj, 894, 43,
  \dodoi{10.3847/1538-4357/ab8799}

\bibitem[{Kann {et~al.}(2023)Kann, Agayeva, Aivazyan, Zhang, Zhao, \& Zhao}]
{Kann_2023_Agayeva_apjl_v948.p12..12L}
Kann, D.~A., Agayeva, S., Aivazyan, V., {et~al.} 2023, \apjl, 948, L12,
  \dodoi{10.3847/2041-8213/acc8d0}

\bibitem[{Kumar {et~al.}(2012)Kumar, Hern{\'a}ndez, Bo{\v{s}}njak, \&
  Barniol~Duran}]{Kumar_2012_Hernandez_mnras_v427.p40..44L}
Kumar, P., Hern{\'a}ndez, R.~A., Bo{\v{s}}njak, {\v{Z}}., \& Barniol~Duran, R.
  2012, \mnras, 427, L40, \dodoi{10.1111/j.1745-3933.2012.01341.x}

\bibitem[{Langer(2012)}]{Langer_2012__araa_v50.p107..164}
Langer, N. 2012, \araa, 50, 107, \dodoi{10.1146/annurev-astro-081811-125534}

\bibitem[{Laskar {et~al.}(2023)Laskar, Alexander, Margutti, Eftekhari,
  Chornock, Berger, Cendes, Duerr, Perley, Ravasio, Yamazaki, Ayache, Barclay,
  Duran, Bhandari, Brethauer, Christy, Coppejans, Duffell, fai Fong, Gomboc,
  Guidorzi, Kennea, Kobayashi, Levan, Lobanov, Metzger, Ros, Schroeder, \&
  Williams}]{Laskar_2023_Alexander_apjl_v946.p23..23L}
Laskar, T., Alexander, K.~D., Margutti, R., {et~al.} 2023, \apjl, 946, L23,
  \dodoi{10.3847/2041-8213/acbfad}

\bibitem[{Lemoine(2013)}]{Lemoine_2013__mnras_v428.p845..866}
Lemoine, M. 2013, \mnras, 428, 845, \dodoi{10.1093/mnras/sts081}

\bibitem[{Lemoine(2015)}]{Lemoine_2015__JournalofPlasmaPhysics_v81.p455810101..455810101}
---. 2015, Journal of Plasma Physics, 81, 455810101,
  \dodoi{10.1017/S0022377814000920}

\bibitem[{Lesage {et~al.}(2023)Lesage, Veres, Briggs, Valverde, Venters, Wadiasingh, Wood, \&
  Zaharijas}]{Lesage_2023_Veres_apjl_v952.p42..42L}
Lesage, S., Veres, P., Briggs, M.~S., {et~al.} 2023, \apjl, 952, L42,
  \dodoi{10.3847/2041-8213/ace5b4}

\bibitem[Levan et al.(2023)]{Levan_2023_Lamb_apj_v946L..p28L}
Levan, A.~J., Lamb, G.~P., Schneider, B., et al.\ 2023, \apjl, 946, L28.
\dodoi{10.3847/2041-8213/acc2c1}

\bibitem[{Lewis(2019)}]{Lewis_2019___v.p..}
Lewis, A. 2019.
\newblock \doarXiv{1910.13970}

\bibitem[{{LHAASO Collaboration} {et~al.}(2023){LHAASO Collaboration}, Cao,
  Aharonian, An, Zhu, Zhu, \&
  Zuo}]{LHAASOCollaboration_2023_Cao_Science_v380.p1390..1396}
{LHAASO Collaboration}, Cao, Z., Aharonian, F., {et~al.} 2023, Science, 380,
  1390, \dodoi{10.1126/science.adg9328}

\bibitem[{Li {et~al.}(2020{\natexlab{a}})Li, Wang, Zheng, Pozanenko,
  Filippenko, Qin, Wang, Jiang, Li, Lin, Liang, Volnova, Elenin, Klunko,
  Inasaridze, Kusakin, \& Lu}]{Li_2020_Wang_apj_v900.p176..176}
Li, L., Wang, X.-G., Zheng, W., {et~al.} 2020{\natexlab{a}}, \apj, 900, 176,
  \dodoi{10.3847/1538-4357/aba757}

\bibitem[{Li {et~al.}(2020{\natexlab{b}})Li, Yang, \&
  Dai}]{Li_2020_Yang_apj_v896.p71..71}
Li, Q.-C., Yang, Y.-P., \& Dai, Z.-G. 2020{\natexlab{b}}, \apj, 896, 71,
  \dodoi{10.3847/1538-4357/ab8db8}

\bibitem[{Liang {et~al.}(2013)Liang, Li, Gao, Zhang, Liang, Wu, Yi, Dai, Tang,
  Chen, Lü, Zhang, Lu, Lü, \& Wei}]{Liang_2013_Li_apj_v774.p13..13}
Liang, E.-W., Li, L., Gao, H., {et~al.} 2013, \apj, 774, 13,
  \dodoi{10.1088/0004-637x/774/1/13}

\bibitem[{Liu {et~al.}(2018)Liu, Wang, Wang, \&
  Dai}]{Liu_2018_Wang_apj_v856.p59..59}
Liu, L.-D., Wang, L.-J., Wang, S.-Q., \& Dai, Z.-G. 2018, \apj, 856, 59,
  \dodoi{10.3847/1538-4357/aab157}

\bibitem[{Lü {et~al.}(2018)Lü, Lan, Zhang, Liang, Kann, Du, \&
  Shen}]{Lue_2018_Lan_apj_v862.p130..130}
Lü, H.-J., Lan, L., Zhang, B., {et~al.} 2018, \apj, 862, 130,
  \dodoi{10.3847/1538-4357/aacd03}

\bibitem[{{MAGIC Collaboration} {et~al.}(2019){MAGIC Collaboration}, {Acciari},
  {Ansoldi}, {Antonelli}, {Engels}, {Baack}, {Babi{\'c}}, {Banerjee}, {Barres
  de Almeida}, {Barrio}, \&
  et~al.}]{MAGICCollaboration_2019_Acciari_nat_v575.p459..463}
{MAGIC Collaboration}, {Acciari}, V.~A., {Ansoldi}, S., {et~al.} 2019, \nat,
  575, 459, \dodoi{10.1038/s41586-019-1754-6}

\bibitem[{Marchenko {et~al.}(2007)Marchenko, Foellmi, Moffat, Martins, Bouret,
  \& Depagne}]{Marchenko_2007_Foellmi_apjl_v656.p77..80L}
Marchenko, S.~V., Foellmi, C., Moffat, A.~F.~J., {et~al.} 2007, \apjl, 656,
  L77, \dodoi{10.1086/512725}

\bibitem[{Margalit \&
  Quataert(2021)}]{Margalit_2021_Quataert_apjl_v923.p14..14L}
Margalit, B., \& Quataert, E. 2021, \apjl, 923, L14,
  \dodoi{10.3847/2041-8213/ac3d97}

\bibitem[{Miceli \& Nava(2022)}]{Miceli_2022_Nava_Galaxies_v10.p66..66}
Miceli, D., \& Nava, L. 2022, Galaxies, 10, 66,
  \dodoi{10.3390/galaxies10030066}

\bibitem[{Milisavljevic {et~al.}(2015)Milisavljevic, Margutti, Kamble,
  Patnaude, Raymond, Eldridge, Fong, Bietenholz, Challis, Chornock, Drout,
  Fransson, Fesen, Grindlay, Kirshner, Lunnan, Mackey, Miller, Parrent,
  Sanders, Soderberg, \&
  Zauderer}]{Milisavljevic_2015_Margutti_apj_v815.p120..120}
Milisavljevic, D., Margutti, R., Kamble, A., {et~al.} 2015, \apj, 815, 120,
  \dodoi{10.1088/0004-637X/815/2/120}

\bibitem[{Mirabal {et~al.}(2003)Mirabal, Halpern, Chornock, Filippenko,
  Terndrup, Armstrong, Kemp, Thorstensen, Tavarez, \&
  Espaillat}]{Mirabal_2003_Halpern_apj_v595.p935..949}
Mirabal, N., Halpern, J.~P., Chornock, R., {et~al.} 2003, \apj, 595, 935,
  \dodoi{10.1086/377471}

\bibitem[{Monfardini {et~al.}(2006)Monfardini, Kobayashi, Guidorzi, Carter,
  Mundell, Bersier, Gomboc, Melandri, Mottram, Smith, \&
  Steele}]{Monfardini_2006_Kobayashi_apj_v648.p1125..1131}
Monfardini, A., Kobayashi, S., Guidorzi, C., {et~al.} 2006, \apj, 648, 1125,
  \dodoi{10.1086/506170}

\bibitem[{Murase {et~al.}(2011)Murase, Toma, Yamazaki, \&
  M{\'{e}}sz{\'{a}}ros}]{Murase_2011_Toma_apj_v732.p77..77}
Murase, K., Toma, K., Yamazaki, R., \& M{\'{e}}sz{\'{a}}ros, P. 2011, \apj,
  732, 77, \dodoi{10.1088/0004-637x/732/2/77}

\bibitem[{Nakar {et~al.}(2009)Nakar, Ando, \&
  Sari}]{Nakar_2009_Ando_apj_v703.p675..691}
Nakar, E., Ando, S., \& Sari, R. 2009, \apj, 703, 675,
  \dodoi{10.1088/0004-637X/703/1/675}

\bibitem[{Nava(2018)}]{Nava_2018__InternationalJournalofModernPhysicsD_v27.p1842003..1842003}
Nava, L. 2018, International Journal of Modern Physics D, 27, 1842003,
  \dodoi{10.1142/S0218271818420038}

\bibitem[{Nava(2021)}]{Nava_2021__Universe_v7.p503..503}
---. 2021, Universe, 7, 503, \dodoi{10.3390/universe7120503}

\bibitem[{Nava {et~al.}(2013)Nava, Sironi, Ghisellini, Celotti, \&
  Ghirlanda}]{Nava_2013_Sironi_mnras_v433.p2107..2121}
Nava, L., Sironi, L., Ghisellini, G., Celotti, A., \& Ghirlanda, G. 2013,
  \mnras, 433, 2107, \dodoi{10.1093/mnras/stt872}

\bibitem[{Nedora {et~al.}(2023)Nedora, Dietrich, Shibata, Pohl, \&
  Crosato~Menegazzi}]{Nedora_2023_Dietrich_mnras_v520.p2727..2746}
Nedora, V., Dietrich, T., Shibata, M., Pohl, M., \& Crosato~Menegazzi, L. 2023,
  \mnras, 520, 2727, \dodoi{10.1093/mnras/stad175}

\bibitem[{O'Connor {et~al.}(2023)O'Connor, Troja, Ryan, Beniamini, van Eerten,
  Granot, Dichiara, Ricci, Lipunov, Gillanders, Gill, Moss, Anand, Andreoni,
  Becerra, Buckley, Butler, Cenko, Chasovnikov, Durbak, Francile, Hammerstein,
  van~der Horst, Kasliwal, Kouveliotou, Kutyrev, Lee, Srinivasaragavan,
  Topolev, Watson, Yang, \&
  Zhirkov}]{OConnor_2023_Troja_ScienceAdvances_v9.p..}
O'Connor, B., Troja, E., Ryan, G., {et~al.} 2023, Science Advances, 9,
  \dodoi{10.1126/sciadv.adi1405}

\bibitem[{{Pe'er}(2012)}]{PeEr_2012__apjl_v752.p8..11}
{Pe'er}, A. 2012, \apjl, 752, 8, \dodoi{10.1088/2041-8205/752/1/L8}

\bibitem[{{Pei}(1992)}]{Pei_1992__apj_v395.p130..130}
{Pei}, Y.~C. 1992, \apj, 395, 130, \dodoi{10.1086/171637}

\bibitem[{Piron(2016)}]{Piron_2016__ComptesRendusPhysique_v17.p617..631}
Piron, F. 2016, Comptes Rendus Physique, 17, 617,
  \dodoi{10.1016/j.crhy.2016.04.005}

\bibitem[{{Planck Collaboration} {et~al.}(2016){Planck Collaboration}, {Ade},
  {Aghanim}, {Arnaud}, {Ashdown}, {Aumont}, {Baccigalupi}, {Banday},
  {Barreiro}, {Bartlett}, \&
  et~al.}]{PlanckCollaboration_2016_Ade_aap_v594.p13..13A}
{Planck Collaboration}, {Ade}, P.~A.~R., {Aghanim}, N., {et~al.} 2016, \aap,
  594, A13, \dodoi{10.1051/0004-6361/201525830}

\bibitem[{Ramirez-Ruiz {et~al.}(2005)Ramirez-Ruiz, Garc{\'\i}a-Segura,
  Salmonson, \&
  P{\'e}rez-Rend{\'o}n}]{RamirezRuiz_2005_GarciaSeg_apj_v631.p435..445}
Ramirez-Ruiz, E., Garc{\'\i}a-Segura, G., Salmonson, J.~D., \&
  P{\'e}rez-Rend{\'o}n, B. 2005, \apj, 631, 435, \dodoi{10.1086/432433}

\bibitem[{Ren {et~al.}(2020)Ren, Lin, Zhang, Wang, Li, Wang, \&
  Liang}]{Ren_2020_Lin_apj_v901.p26..26L}
Ren, J., Lin, D.-B., Zhang, L.-L., {et~al.} 2020, \apj, 901, L26,
  \dodoi{10.3847/2041-8213/abb672}

\bibitem[{Ren {et~al.}(2023)Ren, Wang, Zhang, \&
  Dai}]{Ren_2023_Wang_apj_v947.p53..53}
Ren, J., Wang, Y., Zhang, L.-L., \& Dai, Z.-G. 2023, \apj, 947, 53,
  \dodoi{10.3847/1538-4357/acc57d}

\bibitem[{{Sahu} {et~al.}(2022){Sahu}, {Valadez Polanco}, \&
  {Rajpoot}}]{Sahu_2022_Valadez_ApJ...929...70S}
{Sahu}, S., {Valadez Polanco}, I.~A., \& {Rajpoot}, S. 2022, \apj, 929, 70,
  \dodoi{10.3847/1538-4357/ac5cc6}

\bibitem[{Salafia {et~al.}(2020)Salafia, Barbieri, Ascenzi, \&
  Toffano}]{Salafia_2020_Barbieri_aap_v636.p105..105A}
Salafia, O.~S., Barbieri, C., Ascenzi, S., \& Toffano, M. 2020, \aap, 636,
  A105, \dodoi{10.1051/0004-6361/201936335}

\bibitem[{Salafia {et~al.}(2022)Salafia, Ravasio, Yang, An, Orienti, Ghirlanda,
  Nava, Giroletti, Mohan, Spinelli, Zhang, Marcote, Cim{\`{o}}, Wu, \&
  Li}]{Salafia_2022_Ravasio_apjl_v931.p19..19L}
Salafia, O.~S., Ravasio, M.~E., Yang, J., {et~al.} 2022, \apjl, 931, L19,
  \dodoi{10.3847/2041-8213/ac6c28}

\bibitem[{Saldana-Lopez {et~al.}(2021)Saldana-Lopez, Dom{\'\i}nguez,
  P{\'e}rez-Gonz{\'a}lez, Finke, Ajello, Primack, Paliya, \&
  Desai}]{SaldanaLopez_2021_Dominguez_mnras_v507.p5144..5160}
Saldana-Lopez, A., Dom{\'\i}nguez, A., P{\'e}rez-Gonz{\'a}lez, P.~G., {et~al.}
  2021, \mnras, 507, 5144, \dodoi{10.1093/mnras/stab2393}

\bibitem[{Sari \& Esin(2001)}]{Sari_2001_Esin_apj_v548.p787..799}
Sari, R., \& Esin, A.~A. 2001, \apj, 548, 787, \dodoi{10.1086/319003}

\bibitem[{{Sari} \& {Piran}(1999)}]{Sari-1999-Piran-ApJ...517L.109S}
{Sari}, R., \& {Piran}, T. 1999, \apjl, 517, L109, \dodoi{10.1086/312039}

\bibitem[{Sari {et~al.}(1998)Sari, Piran, \&
  Narayan}]{Sari_1998_Piran_apj_v497.p17..20L}
Sari, R., Piran, T., \& Narayan, R. 1998, \apj, 497, L17,
  \dodoi{10.1086/311269}

\bibitem[{{Sato} {et~al.}(2023){Sato}, {Murase}, {Ohira}, \&
  {Yamazaki}}]{Sato_2023_Murase_mnras_v522.p56..60L}
{Sato}, Y., {Murase}, K., {Ohira}, Y., \& {Yamazaki}, R. 2023, \mnras, 522,
  L56, \dodoi{10.1093/mnrasl/slad038}

\bibitem[{{Schlafly} \&
  {Finkbeiner}(2011)}]{Schlafly_2011_Finkbeiner_apj_v737.p103..103}
{Schlafly}, E.~F., \& {Finkbeiner}, D.~P. 2011, \apj, 737, 103,
  \dodoi{10.1088/0004-637X/737/2/103}

\bibitem[{Schulze {et~al.}(2011)Schulze, Klose, Bj{\"o}rnsson, Jakobsson, Kann,
  Rossi, Kr{\"u}hler, Greiner, \&
  Ferrero}]{Schulze_2011_Klose_aap_v526.p23..23A}
Schulze, S., Klose, S., Bj{\"o}rnsson, G., {et~al.} 2011, \aap, 526, A23,
  \dodoi{10.1051/0004-6361/201015581}

\bibitem[Shrestha et al.(2023)]{Shrestha_2023_Sand_apj_v946L..p25S}
Shrestha, M., Sand, D.~J., Alexander, K.~D., et al.\ 2023, \apjl, 946, L25.
\dodoi{10.3847/2041-8213/acbd50}

\bibitem[Srinivasaragavan et al.(2023)]{Srinivasaragavan_2023_O'Connor_apjl_v949.p39..39S}
Srinivasaragavan, G.~P., O'Connor, B., Cenko, S.~B., et al.\ 2023, \apjl, 949, L39.
\dodoi{10.3847/2041-8213/accf97}

\bibitem[{Suda {et~al.}(2021)Suda, Artero, Asano, Wunderlich, Yamamoto, \&
  Zarić}]{Suda-2022-Artero-icrc.confE.797S}
Suda, Y., Artero, M., Asano, K., {et~al.} 2021, in Proceedings of 37th
  International Cosmic Ray Conference {\textemdash} PoS(ICRC2021), Vol. 395,
  797, \dodoi{10.22323/1.395.0797}

\bibitem[{Tian {et~al.}(2022)Tian, Qin, Du, Yi, \&
  Tang}]{Tian_2022_Qin_apj_v925.p54..54}
Tian, X., Qin, Y., Du, M., Yi, S.-X., \& Tang, Y.-K. 2022, \apj, 925, 54,
  \dodoi{10.3847/1538-4357/ac3de4}

\bibitem[{Tiengo {et~al.}(2023)Tiengo, Pintore, Vaia, Filippi, Sacchi,
  Esposito, Rigoselli, Mereghetti, Salvaterra, {\v{S}}iljeg, Bracco,
  Bo{\v{s}}njak, Jeli{\'c}, \&
  Campana}]{Tiengo_2023_Pintore_apjl_v946.p30..30L}
Tiengo, A., Pintore, F., Vaia, B., {et~al.} 2023, \apjl, 946, L30,
  \dodoi{10.3847/2041-8213/acc1dc}

\bibitem[{Vasilopoulos {et~al.}(2023)Vasilopoulos, Karavola, Stathopoulos, \&
  Petropoulou}]{Vasilopoulos_2023_Karavola_mnras_v521.p1590..1600}
Vasilopoulos, G., Karavola, D., Stathopoulos, S.~I., \& Petropoulou, M. 2023,
  \mnras, 521, 1590, \dodoi{10.1093/mnras/stad375}

\bibitem[{{Veres} {et~al.}(2022){Veres}, {Burns}, {Bissaldi}, {Lesage}, \&
  {Roberts}}]{GCN32636}
{Veres}, P., {Burns}, E., {Bissaldi}, E., {Lesage}, S., \& {Roberts}, o. 2022,
  GRB Coordinates Network, 32636, 1

\bibitem[{Virtanen {et~al.}(2020)Virtanen, Gommers, Oliphant, Haberland, Reddy,
  Cournapeau, Burovski, Peterson, Weckesser, Bright, {van der Walt}, Brett,
  Wilson, Millman, Mayorov, Nelson, Jones, Kern, Larson, Carey, Polat, Feng,
  Moore, {VanderPlas}, Laxalde, Perktold, Cimrman, Henriksen, Quintero, Harris,
  Archibald, Ribeiro, Pedregosa, {van Mulbregt}, \& {SciPy 1.0
  Contributors}}]{Virtanen_2020_Gommers_NatureMethods_v17.p261..272}
Virtanen, P., Gommers, R., Oliphant, T.~E., {et~al.} 2020, Nature Methods, 17,
  261, \dodoi{10.1038/s41592-019-0686-2}

\bibitem[{Wang {et~al.}(2015)Wang, Zhang, Liang, Gao, Li, Deng, Qin, Tang,
  Kann, Ryde, \& Kumar}]{Wang_2015_Zhang_apjs_v219.p9..9}
Wang, X.-G., Zhang, B., Liang, E.-W., {et~al.} 2015, \apjs, 219, 9,
  \dodoi{10.1088/0067-0049/219/1/9}

\bibitem[{Warren {et~al.}(2022)Warren, Dainotti, Barkov, Ahlgren, Ito, \&
  Nagataki}]{Warren_2022_Dainotti_apj_v924.p40..40}
Warren, D.~C., Dainotti, M., Barkov, M.~V., {et~al.} 2022, \apj, 924, 40,
  \dodoi{10.3847/1538-4357/ac2f43}

\bibitem[{{Waxman}(1997)}]{Waxman_1997_ApJ...485L...5W}
{Waxman}, E. 1997, \apjl, 485, L5, \dodoi{10.1086/310809}

\bibitem[{Woosley \& Bloom(2006)}]{Woosley_2006_Bloom_araa_v44.p507..556}
Woosley, S.~E., \& Bloom, J.~S. 2006, \araa, 44, 507,
  \dodoi{10.1146/annurev.astro.43.072103.150558}

\bibitem[{Yan {et~al.}(2007)Yan, Wei, \&
  Fan}]{Yan_2007_Wei_ChineseJournalofAstronomyandAstrophysics_v7.p777..788}
Yan, T., Wei, D.-M., \& Fan, Y.-Z. 2007, Chinese Journal of Astronomy and
  Astrophysics, 7, 777, \dodoi{10.1088/1009-9271/7/6/05}

\bibitem[{Yang {et~al.}(2018)Yang, Zou, Chen, Liao, Lei, \&
  Liu}]{Yang_2018_Zou_ResearchinAstronomyandAstrophysics_v18.p18..18}
Yang, C., Zou, Y.-C., Chen, W., {et~al.} 2018, Research in Astronomy and
  Astrophysics, 18, 018, \dodoi{10.1088/1674-4527/18/2/18}

\bibitem[{Yang {et~al.}(2023)Yang, Zhao, Yan, Wang, Zhang, An, Cai, Li, Li,
  Liu, Liu, Ma, Meng, Peng, Qiao, Shao, Song, Tan, Wang, Wang, Wen, Xiao, Xue,
  Yang, Yin, Zhang, Zhang, Zhang, Zhang, Zheng, Zheng, Xiong, \&
  Zhang}]{Yang_2023_Zhao_apjl_v947.p11..11L}
Yang, J., Zhao, X.-H., Yan, Z., {et~al.} 2023, \apjl, 947, L11,
  \dodoi{10.3847/2041-8213/acc84b}

\bibitem[{Yang {et~al.}(2016)Yang, Zhang, \&
  Dai}]{Yang_2016_Zhang_apjl_v819.p12..12L}
Yang, Y.-P., Zhang, B., \& Dai, Z.-G. 2016, \apjl, 819, L12,
  \dodoi{10.3847/2041-8205/819/1/L12}

\bibitem[{Yi {et~al.}(2013)Yi, Wu, \& Dai}]{Yi_2013_Wu_apj_v776.p120..120}
Yi, S.-X., Wu, X.-F., \& Dai, Z.-G. 2013, \apj, 776, 120,
  \dodoi{10.1088/0004-637X/776/2/120}

\bibitem[{Yi {et~al.}(2020)Yi, Wu, Zou, \& Dai}]{Yi_2020_Wu_apj_v895.p94..94}
Yi, S.-X., Wu, X.-F., Zou, Y.-C., \& Dai, Z.-G. 2020, \apj, 895, 94,
  \dodoi{10.3847/1538-4357/ab8a53}

\bibitem[{Zhang(2018)}]{Zhang_2018pgrb.book.....Z}
Zhang, B. 2018, The Physics of Gamma-Ray Bursts (Cambridge University Press),
  \dodoi{10.1017/9781139226530}

\bibitem[{Zhang \& Kobayashi(2005)}]{Zhang_2005_Kobayashi_apj_v628.p315..334}
Zhang, B., \& Kobayashi, S. 2005, \apj, 628, 315, \dodoi{10.1086/429787}

\bibitem[{Zhang {et~al.}(2003)Zhang, Kobayashi, \&
  M{\'e}sz{\'a}ros}]{Zhang_2003_Kobayashi_apj_v595.p950..954}
Zhang, B., Kobayashi, S., \& M{\'e}sz{\'a}ros, P. 2003, \apj, 595, 950,
  \dodoi{10.1086/377363}

\bibitem[{{Zhang} \&
  {M{\'e}sz{\'a}ros}(2001)}]{Zhang_2001_Meszaros_apj_v559.p110..122}
{Zhang}, B., \& {M{\'e}sz{\'a}ros}, P. 2001, \apj, 559, 110,
  \dodoi{10.1086/322400}

\bibitem[Zhang et al.(2023{\natexlab{a}})]{Zhang_2023_Huang_apj_v956L..p21Z}
Zhang, H.-M., Huang, Y.-Y., Liu, R.-Y., et al.\ 2023, \apjl, 956, L21.
\dodoi{10.3847/2041-8213/acfcab}

\bibitem[{{Zhang} {et~al.}(2021){Zhang}, {Ren}, {Huang}, {Liang}, {Lin}, \&
  {Liang}}]{Zhang_2021_Ren_apj_v917.p95..95}
{Zhang}, L.-L., {Ren}, J., {Huang}, X.-L., {et~al.} 2021, \apj, 917, 95,
  \dodoi{10.3847/1538-4357/ac0c7f}

\bibitem[{Zhang {et~al.}(2023{\natexlab{b}})Zhang, Ren, Wang, \&
  Liang}]{Zhang_2023_Ren_apj_v952.p127..127}
Zhang, L.-L., Ren, J., Wang, Y., \& Liang, E.-W. 2023{\natexlab{b}}, \apj, 952,
  127, \dodoi{10.3847/1538-4357/acd190}

\bibitem[{{Zhang} {et~al.}(2022){Zhang}, {Xin}, {Wang}, {Han}, {Xu}, {Zhu},
  {Wu}, {Wei}, \& {Liang}}]{Zhang_2022_Xin_apj_v941.p63..63}
{Zhang}, L.-L., {Xin}, L.-P., {Wang}, J., {et~al.} 2022, \apj, 941, 63,
  \dodoi{10.3847/1538-4357/aca08f}

\bibitem[Zou et al.(2009)]{Zou_2009_Fan_mnras_v396.p1163Z}
 Zou, Y.-C., Fan, Y.-Z., \& Piran, T.\ 2009, \mnras, 396, 1163.
 \dodoi{10.1111/j.1365-2966.2009.14779.x}
\end{thebibliography}

\end{document}